\documentclass{article}
\usepackage{emulateapj}
\usepackage{apjfonts}

\usepackage{graphics}

\newcommand{\gs}{\mathrel{\mathchoice{\vcenter{\offinterlineskip\halign{\hfil
$\displaystyle##$\hfil\cr>\cr\sim\cr}}}
{\vcenter{\offinterlineskip\halign{\hfil$\textstyle##$\hfil\cr>\cr\sim\cr}}}
{\vcenter{\offinterlineskip\halign{\hfil$\scriptstyle##$\hfil\cr>\cr\sim\cr}}}
{\vcenter{\offinterlineskip\halign{\hfil$\scriptscriptstyle##$\hfil\cr>\cr\sim
\cr}}}}}

\newcommand{\lya}{Ly$\alpha$}
\newcommand{\lyb}{Ly$\beta$}
\newcommand{\lyg}{Ly$\gamma$}
\newcommand{\HI}{\ion{H}{1}}
\newcommand{\CIV}{\ion{C}{4}}
\newcommand{\SiIV}{\ion{Si}{4}}
\newcommand{\MgII}{\ion{Mg}{2}}
\newcommand{\kms}{km s$^{-1}$} 
\def\ltsima{$\; \buildrel < \over \sim \;$}
\def\simlt{\lower.5ex\hbox{\ltsima}}
\def\gtsima{$\; \buildrel > \over \sim \;$}
\def\simgt{\lower.5ex\hbox{\gtsima}}

\lefthead{Measuring the \CIV/\HI\ Ratio in the \lya\ Forest.}
\righthead{Ellison et al.}

\begin{document}


\submitted{Accepted for publication in the Astronomical Journal}

\title{The Enrichment History of the Intergalactic Medium - \\
Measuring the \CIV/\HI\ Ratio in the \lya\ Forest\altaffilmark{1}.}


\author{Sara L. Ellison\altaffilmark{2, 3}, 
Antoinette Songaila\altaffilmark{4}, Joop Schaye\altaffilmark{3},
Max Pettini\altaffilmark{3}}

\altaffiltext{1}{The data
presented herein were obtained at the W. M. Keck Observatory, which is
operated as a scientific partnership among the California Institute of
Technology, the University of California and the National Aeronautics
and Space Administration.  The Observatory was made possible by the
generous financial support of the W. M. Keck Foundation.}
\altaffiltext{2}{sara@ast.cam.ac.uk}
\altaffiltext{3}{Institute of Astronomy, Madingley Road, Cambridge CB3 0HA, UK}
\altaffiltext{4}{Institute for Astronomy, 2680 Woodlawn Drive,
University of Hawaii, Honolulu, HI 96822, USA}



\begin{abstract}

We have obtained an exceptionally high S/N, high resolution spectrum
of the gravitationally lensed quasar Q1422+231.  A total of 34
\CIV\ systems are identified, several of which had not been seen in
previous spectra.  Voigt profiles are fitted to these
\CIV\ systems and to the entire \lya\ forest in order to determine
column densities, $b$-values and redshifts for each absorption component.
The column density distribution for \CIV\ is found to be a
power law with index $\alpha = 1.44 \pm 0.05$, down to at least log $N$(\CIV)
= 12.3. We use simulations to estimate the incompleteness correction and
find that there is in fact no evidence for flattening of the power
law down to log $N$(\CIV) =
11.7 -- a factor of ten lower than previous measurements. In order to
determine whether the \CIV\ enrichment extends to even lower column
density \HI\ clouds,  we utilize two analysis
techniques to probe the low column density regime in the \lya\
forest.  Firstly, a composite stacked spectrum is produced by
combining the data for Q1422+231 and another bright QSO, APM
08279+5255.   The S/N of the stacked spectrum is 1250 and yet no resultant
\CIV\ absorption is detected.  We discuss the various
problems that affect the stacking technique and focus in particular on
a random velocity offset between \HI\ and its associated \CIV\ which we
measure to have a dispersion of $\sigma_v$ = 17 \kms.  
It is concluded that, in our data, this offset results in an
underestimate of the amount of \CIV\ present
by a factor of about two and this technique is therefore not sufficiently 
sensitive to probe the low column density \lya\ clouds to meaningful
metallicities.  Secondly, we use measurements of individual pixel
optical depths of \lya\ and corresponding \CIV\ lines.  We compare the
results obtained from this optical depth method with analyses of
simulated spectra enriched with varying \CIV\ enrichment recipes.  From these
simulations, we conclude that more \CIV\ than is currently directly
detected in Q1422+231 is required to reproduce the optical depths
determined from the data, consistent with the conclusions drawn from
consideration of the power law distribution.

\end{abstract}




\section{Introduction}

Our understanding of the intergalactic medium (IGM) has undergone a paradigm
shift in recent years, largely as a result of powerful hydrodynamical
simulations.  When the \lya\ forest was first observed in the late 1960s, 
the rich field of \HI\ absorption blueward of the QSO's \lya\ emission
was interpreted as the detection of discrete intergalactic clouds
(e.g.\ Lynds \& Stockton 1966; Lynds 1971; Sargent et al.\ 1980).  In
order to understand the existence of such isolated absorbers, various
theories of cloud confinement, including self-gravity (Melott 1980),
cold dark matter minihaloes (Rees 1986) and the presence
of an inter-cloud medium (e.g.\ Sargent et al.\ 1980;
Ostriker \& Ikeuchi 1983) were proposed.   However, the advent of
hydrodynamical simulations, which model the growth of structure in the
high redshift universe, provided a significant revision to our picture of the
IGM (see the recent review by Efstathiou, Schaye and Theuns 2000).  
It has been found that in the presence of a UV ionizing
background, the `bottom-up' hierarchy of structure formation knitted a
complex, but smoothly fluctuating `cosmic web' in the
IGM (e.g.\ Cen et al.\ 1994; Hernquist et al.\ 1996; Bi \& Davidsen 1997).  
The absorption in the \lya\ forest is caused not by individual,
confined clouds, but by a gradually varying density field
characterized by overdense sheets and filaments and extensive,
underdense voids.
The advance in theoretical simulations has been matched by increasingly
high quality data, as comprehensively reviewed by Rauch (1998).  One of the
major discoveries concerning the IGM has been the 
identification of metal absorption lines associated with many of the
\lya\ forest clouds (Cowie et al.\ 1995; Tytler et al.\ 1995). 
Thus, whilst the \lya\ forest was once thought to
be chemically pristine, it has now been well-established that a large
fraction of the high column density \lya\ clouds ($N$(\HI) $>$ 14.5), 
associated with
collapsing, over-dense structures, contain metals (most notably \CIV)
-- the signature of enrichment by the products of stellar nucleosynthesis
(Songaila \& Cowie 1996).  

The presence of metals in the \lya\ forest may be reasonably explained
either by
in-situ enrichment (local star formation in the \HI\ cloud itself
or in a nearby galaxy)  or by
early pre-enrichment by a high redshift episode of Population III
stars.   Whilst the effects of
supernova feedback are still not fully understood and therefore the
spatial extent of wind-driven ejecta is poorly constrained,
enrichment by galactic winds and superbubbles is
unlikely to be an efficient way to distribute metals over
distances large in comparison with the mean separation between galaxies 
(MacLow \& Ferrara 1999).  
Instead, models have appealed to larger scale processes such as 
merging and turbulent diffusion as the dominant mixing mechanisms
(Gnedin \& Ostriker 1997; Gnedin 1998; Ferrara, Pettini \& 
Shchekinov 2000).  Such processes 
would take time to smooth out the metallicity of the IGM so that at
$z \sim 3$ the metal enrichment is still expected to be very patchy.
Whilst the deep potential wells of galaxies inhibit efficient,
widespread distribution of metals far from their sites of formations,
small regions of star formation at high redshift 
may be able to eject their
nucleosynthetic products for more homogeneous mixing (e.g.\ Nath \& Trentham
1997 and references therein).  An episode of Population III star 
formation may therefore have spread metals far from their 
sites of formation, seeding `sterile' regions of the IGM with metals
(Ostriker \& Gnedin 1996).
Clearly, distinguishing between in-situ and Pop III scenarios has important
implications for understanding not only the first generation of stars,
but also the mechanisms by which metals are mixed and distributed from
their stellar birthplaces.

Several papers (Cowie \& Songaila 1998; Lu et al.\ 1998; Ellison et al.\ 1999a,
hereafter Paper I) have addressed this question by attempting to probe the low
density regions of the IGM where the difference in metallicity
predicted by in-situ and 
Pop III enrichment may be most marked.  The detection of \CIV\ 
$\lambda\lambda$1548,1550 associated with low column density 
\lya\ clouds (log $N$(\HI) $<$ 
14.0) is observationally challenging, even with the capabilities of Keck, due
to the extreme weakness of the metal lines.  Analysis techniques have
therefore been developed to effectively enhance the sensitivity of the data
beyond the normal equivalent width limits of the spectra. 
Two in particular
have received recent attention, namely the production of a stacked \CIV\
spectrum (Lu et al.\ 1998) and the use of individual pixel optical depths of
Ly~$\alpha$\ and \CIV\ (Cowie \& Songaila 1998). 
In Paper I we attempted to
reconcile the apparently conflicting results obtained from these two
techniques with an analysis of a very high signal-to-noise ratio (S/N)
spectrum of the ultra-luminous BAL quasar APM 08279+5255.  Rigorous testing of
the analysis procedures revealed that both methods suffered from hitherto
unrecognised limitations and it was concluded that the question of whether or
not the low density regions of the IGM have been enriched remains unanswered.

In order to probe deeper into the low density IGM, we have obtained an
exceptionally high S/N spectrum of the well-known lensed QSO, Q1422+231, using
the HIRES instrument  (Vogt 1994) on the KeckI telescope.  Although this QSO
has been well studied in the past (e.g.\ Songaila \& Cowie 1996;  Songaila
1998), 
we have roughly doubled the exposure time of earlier spectra, allowing us 
to probe the metallicity of the \lya\
forest to more sensitive levels than has previously been achieved.  
The spectrum is much better suited to the present work than the the much more
complex BAL quasar APM 08279+5255 (Ellison et al.\ 1999a).
We present a
careful and extensive analysis of the \CIV\ systems in order to determine the
extent of metal enrichment in the IGM.  This paper is organised as follows.
In \S2 we describe the observations, the data reduction procedures and the
Voigt profile fitting process used to determine column densities.  We briefly
discuss in \S3 the suitability of Q1422+231 for this work and define the
redshift interval over which we will perform the analysis.  In \S4, we
determine the column densities of the 34 detected \CIV\ absorption systems in
the spectrum and investigate the column density distribution of these
absorbers.  Finally, we critically assess the two methods of analysis,
described in sections \ref{stacksec} and \ref{tausec}, which have been
developed to probe low column density \lya\ clouds to very sensitive levels.
We utilize a suite of simulation techniques to fully test these methods and
quantify the potential inaccuracies in our analysis.

We adopt $\Omega_M$ = 1.0 throughout.

\section{Data Acquisition and Reduction}

Observations of Q1422+231 were made with the HIRES spectrograph (Vogt 1994) on
the KeckI telescope in February 1999, using the $1.14 \times 7$\ arcsec slit.
The resultant resolution is $R = 37,000$.  Individual exposure times varied
between 30 min and 40 min, for a total exposure of 630 min in two
(overlapping) grating settings to provide total wavelength coverage between
the quasar's \lya\ and \CIV\ emission lines.  The data were reduced as
described in Songaila (1998) and added to the spectrum of Q1422+231
obtained under similar conditions by Songaila and Cowie (1996).  The
total integration time for the two datasets is 1130 minutes and
has a S/N of 200 -- 300 redward of the QSO's
\lya\ emission, a quality superior to all previously published data.

After extraction and sky subtraction, we noted that the cores of the 
saturated absorption lines contain a very small systematic residual flux.  
This error in the zero level is possibly due to light from a foreground source
or a small underestimate of the background sky.  
The correction required to bring the line cores to zero was found to vary 
slightly with wavelength, ranging from 0.8\% for the bluest absorption
lines ($\lambda < 4850$ \AA) to 0.2\% at $\lambda \sim 5500$ \AA.  No
correction could be estimated redward of the QSO's \lya\ emission
since the absence of saturated lines gave us no basis for estimating
the adjustment
required.  However, since the correction factor required appears to
diminish with increasing wavelength and is already very small at 5500
\AA, and given that such an adjustment will make little difference to the
relatively weak \CIV\ lines studied in the red, we consider any
residual eror to be unimportant for the analysis presented here.

The continuum fit was
achieved using the STARLINK package DIPSO with a cubic spline
polynomial applied to windows of spectrum judged to be free from
absorption.
Voigt profiles were fitted to the entire normalized \lya\ forest using
VPFIT (Webb 1987) to decompose the complex absorption into individual
components defined by a column density ($N$(\HI)), a $b$-value (Doppler width)
and a redshift.  This task,
though time consuming, is an important feature of our analysis since
synthetic spectra can be simulated based on the line list of Voigt
profile parameters.  For example, additional \CIV\ associated with the
fitted \HI\ 
can be included to these synthetic spectra according to any desired 
enrichment recipe.  As will be
discussed later in this paper, this is an essential step for
testing the analysis techniques used here. 

\section{Q1422+231 -- Suitability and Sample Definition}

Q1422+231 ($z_{em} = 3.625$)
is a well-studied quasar and actually consists of four closely
spaced lensed images with separations of 0.5 -- 1.3 arcsec; the
lensing galaxy is at $z_{lens}$= 0.34 (Patnaik et al.\ 1992; Kundic et al
1997; Tonry 1998).  Gravitational lenses enhance the
emission from high redshift QSOs making them more powerful probes of the
intergalactic and interstellar medium, but obviously the sight lines
will sample different spatial regions of the intervening material. 
Its redshift and luminosity (V=16.5) have made  Q1422+231 an ideal 
candidate with which to probe intervening material through closely
spaced sightlines
(Petry et al.\ 1998; Rauch et al.\ 1999). 
The observations reported here are of the closely spaced A and B components.
In our observations the
sightlines are unresolved.    It is well
established that multiple lines of sight through quasar pairs
separated by several arcsecs show coherence between \lya\ clouds
on scales $>$ 100 kpc
(e.g.\ Bechtold et al.\ 1994; Dinshaw et al.\ 1997).  This is consistent
with the scenario that has emerged from hydrodynamical simulations
that portrays structure in the IGM not as discrete localized clouds but
as a smoothly fluctuating medium.  For metal line systems, 
whilst there may be slight differences in the individual components
that constitute \CIV\ complexes, Rauch et al.\ (1998) have shown that the
total system column density remains largely unchanged over $\sim$ 10
kpc.  The small linear separations probed by Q1422+231 ($< 0.14$ kpc
h$^{-1}$ for the \CIV\ systems in the $z$ range considered here)
are therefore unlikely to give rise to by
line of sight differences so large as to compromise our column density
determinations.  

In order to avoid confusion between \lya\ and
higher order Lyman lines, we restrict
our analysis of the \lya\ forest in Q1422+231 to the interval 4740 
$< \lambda $(\lya)$< 5520 $ \AA\ which corresponds to a redshift range of
2.90 $< z_{abs} < $ 3.54. The lower limit of this interval is determined by the
onset of \lyb\ absorption and the upper limit is enforced in order to
avoid effects due to quasar proximity (e.g.\ Lu, Wolfe \& Turnshek 1991) and
corresponds to a velocity separation of $\sim 5500$ \kms\ relative to
the emission redshift, $z_{em}$ = 3.625.
In order to improve the statistics and S/N of stacked data (\S5), we have,
in some of the work presented here, included the spectrum of APM
08279+5255, recently analysed in Paper I.  APM 08279+5255
is an ultra-luminous Broad Absorption Line (BAL) quasar with a
systemic redshift $z_{em}$ = 3.911 and a broad band magnitude R =
15.2.  The data obtained for this quasar (as presented in Paper I)
are also of excellent quality,
though the broad absorption line makes some portions of the spectrum less
ideal for this problem than the spectrum of Q1422+231,
and taken together with
Q1422+231 they represent a premium data set for this work.  The
wavelength region used for APM 08279+5255 is 4995 $< \lambda $(\lya)$<
5720$ \AA\ (3.11 $<z_{abs}<$ 3.70) which in addition to the selection criteria 
applied to Q1422+231, takes into account the BAL nature of this
quasar. Finally, a region contaminated by atmospheric absorption
from 6865 $< \lambda $(\CIV) $<$ 6940 \AA\ (corresponding to 
$3.43 < z < 3.48$\ for \CIV) was excluded in both spectra. 

\section{Detected \CIV\ Systems}

Q1422+231 has been the target of several observing campaigns over
the years, which have resulted in spectra of various S/N ratios.    
Whilst the primary motivation for obtaining such a high S/N spectrum was our
scientific goal of probing the low column density \lya\ forest,
continued focus on this target has produced important results for IGM
enrichment on a range of column density scales.  In this section we
present all of the 
detected \CIV\ systems within the redshift range defined in \S3,
several of which have not been detected in previous spectra and we show
that the power law column density distribution established for strong
absorbers with $N$(\CIV)$>$ 12.75  (Songaila 1997) 
continues to significantly lower column densities.

We detect 34 \CIV\ systems within our defined wavelength interval, 29 of
which are associated with saturated \lya\ clouds  (log $N$(\HI) $\gs$ 14.5).
These systems exhibit a wide variety of column densities and
complexity (i.e.\ number of components) as can be seen in Figure
\ref{civsys}.  The column densities, $b$-values and redshifts
determined for each \CIV\ component using the Voigt profile fitting program
VPFIT, are presented in Table \ref{civlist}\footnote{The 1548 \AA\
doublet component of system C11 is blended 
with the 1550 \AA\ component of C10.  Both of these components fall at
6302 \AA\ which coincides with a strong sky line,
whose subtraction has left some residuals in our spectrum.  However,
the C11 system is also observed in the spectrum of Boksenberg, Sargent
and Rauch 
(in preparation) where clearly more absorption is required at 6302 \AA\ than is
accounted for by the 1550 \AA\ line of C10.  
The Voigt profile for C11 presented here is based only on the 
1550 \AA\ line which itself is found in a region of weak atmospheric
absorption and therefore is likely to be less reliable than our other
fits}.  
Examples of \CIV\ systems detected in our
spectrum which have not been observed in other published spectra
(Songaila \& Cowie 1996; Songaila 1998) 
are the weak systems at $z_{abs} = 3.276, 3.317$ and 3.518 (C24, C25
and C33 in Figure \ref{civsys}).

\subsection{Column Density Distribution of \CIV\ Systems}

\begin{figure}
\centerline{\rotatebox{270}{\resizebox{0.7\textwidth}{!}
{\includegraphics{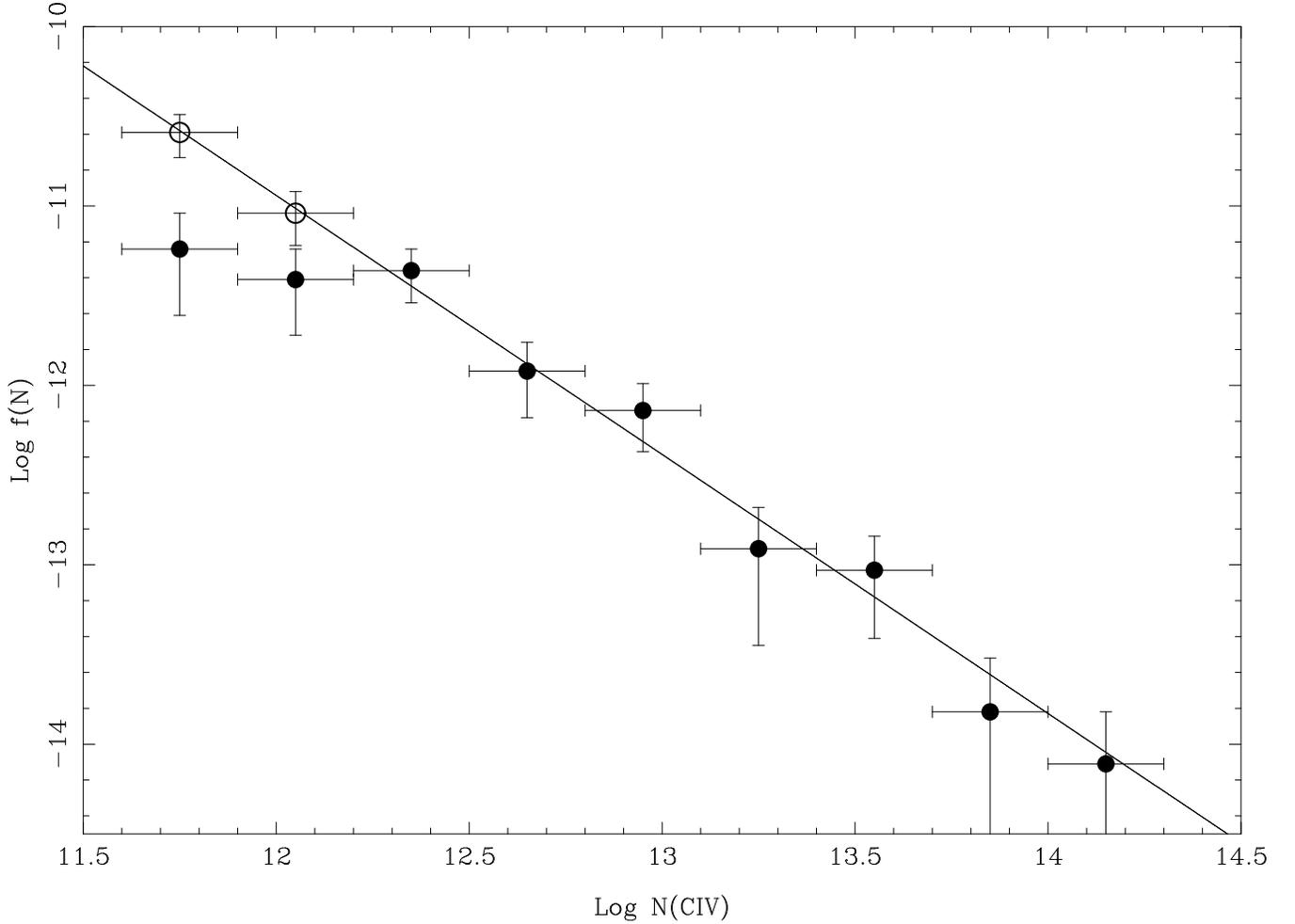}}}}
\caption{Column density distribution of C\,IV absorbers in
Q1422+231 where $f$(N) is the number of systems per column density
interval per unit redshift path. 
The data (solid points) have been grouped into bins of width 0.3
in log~$N$(C\,IV) for display purposes only.   A maximum likelihood
analysis of the distribution, assumed to be a power-law of the
form $f(N) dN = BN^{-\alpha} dN$, returned a best fit index
$\alpha = 1.44 \pm 0.05$ indicated by the solid line.  At low
$N$(C\,IV), lines with large $b$-values may not have been identified
in our spectrum.  We therefore simulated synthetic absorption lines for the two
lowest column density bins shown here, both of which show signs of
incompleteness.   A total of 40 lines with $b$-values drawn at random
for the observed distribution were simulated for each bin
and a correction factor calculated for $N$(C\,IV) = 11.75 ($ \times
4.4$) and $N$(C\,IV) = 12.05 ($\times 2.3$), based on the number 
of unidentifiable lines in each column density simulation. Corrected
points are shown with open circles. }
\label{f_N}
\end{figure}

Previous studies of \CIV\ absorbers in a variety of quasar sightlines has
established a power law column density distribution (with index $\alpha \sim
1.5$, see eqn. 1) complete down to log $N$(\CIV) $\simeq$ 12.75 at $z > 3$ 
and 12.25 at $z < 3$.(Songaila 1997).  
The column density distribution function is defined as

\begin{equation}
f(N)dN = BN^{-\alpha} dN
\end{equation}

where, $f(N)$ is the number of systems per column density interval per
unit redshift path.  The redshift path (used instead of $z$ in order
to account for co-moving distances) is given by $X(z) =
\frac{3}{2}[(1+z)^{\frac{3}{2}} - 1]$ for our adopted cosmology.
An important question is how much of the iceberg has been exposed?  To
what limit does this power law distribution continue? 
As spectral data have improved, early determinations of the
column density distribution of \CIV\ absorbers have been shown to be
incomplete at low column densities (compare for example Petitjean \&
Bergeron 1994 and Songaila 1997).
The superb quality of this single spectrum can address
whether the established power law continues to lower $N$(\CIV), in
which case the apparent fall-off towards low $N$(\CIV) seen previously
is due to incompleteness, or whether there is a real turnover in number
density.  In Figure \ref{f_N} we show the column density distribution
of \CIV\ derived from the systems in Q1422+231, assumed to be a power
law of the form of eqn. 1.  A maximum likelihood fit to
the data (binned in Figure \ref{f_N} for display purposes only)
gives a power law index of $\alpha = 1.44 \pm 0.05$, 
consistent with other recent
estimates.  This high quality spectrum, however, clearly uncovers more
of the `iceberg' than previous studies and the power law continues
down to log $N$(\CIV) $\sim$ 12.3 (Figure \ref{f_N}, solid points). 
Below this column density, $f(N)$ shows an apparent departure from the
power law which may
be due to incompleteness or alternatively reflect
a real turnover in the $N$(\CIV)
distribution.  The formal 5$\sigma$ detection limit for \CIV\
in our spectrum is log $N$(\CIV) = 11.6, for the median $b$-value of 
the observed \CIV\ absorption lines, $b_{median} = 13$ \kms. 
This detection limit is for \textit{one \CIV\
line} and is based on the $\lambda$1548 \AA\ doublet
component.  However,
identification of a suspected \CIV\ system is dependent on confirmation
from the weaker \CIV\ $\lambda$1550 \AA\ line whose oscillator
strength is only half of the $f$-value of the 1548 \AA\ line.  This
effectively reduces our sensitivity for detecting \CIV\
\textit{systems} by a factor of two to a 5$\sigma$ detection
limit of log $N$(\CIV) = 11.9.  Moreover, lines with $b$-values
significantly larger than the median $b$=13 \kms, may not be detected.
In order to estimate the incompleteness for systems with log $N$(\CIV) $<$
12.3 due to large $b$, 40 \CIV\ doublets were simulated for the two
bins in Figure \ref{f_N} which show a departure from the established
power law, i.e.\ 
$N$(\CIV) = 12.05 and $N$(\CIV) = 11.75.  The $b$-value for each line
was drawn at random from the real distribution of Doppler widths and
noise was added at the appropriate level.  Each simulated line was then
inspected to assess whether it would have been identified in the
original spectrum so that a correction factor could be determined to
estimate incompleteness.  For $N$(\CIV) = 12.05 a correction factor of
2.4 was determined (17/40 test \CIV\ systems detected in
the incompleteness trial), with the largest $b$-value in the lines detected
being 13 \kms.  In the lowest column density bin,
$N$(\CIV) = 11.75 only 9 out of 40 \CIV\ lines (with $b <$ 8 \kms)
were identified,
corresponding to a correction factor of 4.4.  Clearly, these
correction factors assume that the $b$-value distribution does not
significantly change with decreasing \CIV\ column density. 
The two data points adjusted for incompleteness
are represented by open circles in Figure \ref{f_N}.  Thus, it appears that
the power law distribution of column densities continues down to at
least log $N$(\CIV) = 11.75.  

\subsection{Homogeneous or Variable Metallicity?}

Recently, hydrodynamical simulations have been used to predict 
the expected scatter in individual \CIV/\HI\ ratios within strong \lya\ 
absorbers for a fixed [C/H] (Hellsten et al.\ 1997; 
Rauch, Haehnelt \& Steinmetz 1997, Dav\'{e} et al.\ 1998).  When
compared with observations, these models show that the data are
consistent with a mean IGM metallicity [C/H] = $-2.5$ with variations
of up to a factor of 10 around this average value.
In these models, the \CIV\
systems associated with high column density \lya\ clouds are 
dominated by metals from in-situ star formation, since
simulations show that these metal enriched clouds are found within a few
tens of kpc from collapsed, dense clumps at $z \sim 3$ 
(e.g.\ Haehnelt 1998 and references therein).  We would then
expect the metallicity of such clouds to be variable, being
dependent on the local star formation history, and therefore the
scatter amongst individual \CIV/\HI\ ratios at a given $N$(\HI) 
to be larger than that predicted for a homogeneous [C/H]. 
 
We can calculate \CIV/\HI\ ratios for 19 systems in our spectrum of
Q1422+231. Of the 34 detected \CIV\ systems in the spectrum presented here, 29
are associated with saturated \lya\ lines.  Although it is not
possible to determine 
accurate \HI\ column densities from these lines alone, for 14 of the
\CIV\ systems higher order Lyman lines are both accessible and
suitably uncontaminated by blends so that an accurate $N$(\HI) can in
fact be
derived.  In addition, 5 of the systems in Q1422+231 are associated with
clouds with log $N$(\HI) $<$ 14.5 which are not saturated so that the
column densities can be determined from the Voigt profile fit of \lya\
only.  The \CIV\ and \HI\ 
column densities determined for these 19 systems are presented in Table
\ref{hifits}.  

Dav\'{e} et al.\ (1998) have also extensively studied the \CIV/\HI\ ratios of
Q1422+231, as determined from the spectrum of Songaila \&
Cowie (1996).  They constructed a mock spectrum of Q1422+231 from
hydrodynamic simulations for comparison with the data.  Having both
simulated and observed spectra at their disposal, they measured
\CIV/\HI\ ratios in both datasets in a \textit{consistent manner}
(using the AutoVP Voigt profile fitter, Dav\'{e} et al.\ 1997), 
and found that an intrinsic scatter of approximately 0.5 dex in
metallicity is required to fit the data.  
However, the observed range of \CIV/\HI\ ratios is the
result of many complex effects, which include not only spatial
variations in the temperature-density relationship, but also, for example, 
fluctuations in the ionizing background. These complex effects have 
not yet been fully incorporated into models and therefore we should 
be mindful that the scatter in measured values of
\CIV/\HI\ is an upper limit to the variations in [C/H] when compared
with homogeneously enriched simulations with a uniform ionizing
background.  

Dav\'{e} et al.\ (1998) also found that the most robust diagnostic
for determining the mean carbon abundance in detected \CIV\ systems is
$\langle$log $N$(\CIV)$\rangle$, although this statistic is clearly
dependent on 
the sensitivity of the data.  As a consistency check, we determine the
$\langle$log $N$(\CIV)$\rangle$ for those \CIV\ systems identified in
our spectrum of 
Q1422+231 whose column densities are above the detection limit of the 
Songaila \& Cowie (1996) spectrum (estimated to be log $N$(\CIV\ $>$ 12.0).  We
calculate $\langle$log $N$(\CIV)$\rangle$ = 12.77 which compares well with the
value of 12.72 determined by Dav\'{e} et al.\ (1998), considering the
errors associated with column density determinations of individual
\CIV\ systems.  This demonstrates that the different
line finding and fitting procedures 
used on different spectra reproduce the same
answer, even when the same systems are observed with a higher
S/N.  For Voigt profile fitting, this is an important point to make in a 
process that is sometimes not clear-cut.  It is also an illustration
that the detection limit of the spectrum used by Dav\'{e} et al.\ (1998)
was well determined and did not suffer from serious incompleteness.
Ideally, one would like to see how the mean [C/H] varies as we detect
progressively weaker \CIV\ systems.  However,
a simple extrapolation to lower column densities is not
possible, since the relation between $\langle$log $N$(\CIV)$\rangle$  and [C/H]
presented in Dav\'{e} et al.\ (1998) is tailored specifically for the
detection limits of their data.  In order to apply this technique to the full
sample of \CIV\ systems detected in the new Q1422+231 spectrum
presented here, it would be necessary to re-create a model spectrum 
to match our data.  However, we remind the reader 
that the caveats mentioned in
relation to the scatter of \CIV/\HI\ are also applicable here, a fact
of which one should be aware when comparing observational and
simulated results and interpreting the conversion to metallicity. 
   
\medskip

We can summarise the results from this section with the following
conclusions.  By obtaining an ultra-high S/N spectrum of the $z =
3.625$ quasar Q1422+231, we have detected 34 \CIV\ absorption systems
with 2.91 $< z_{abs} < $ 3.54.  Some hitherto undetected weak \CIV\
systems are reported whose column densities are consistent with the
established $f(N)$ distribution showing that this power law 
($\alpha = 1.44 \pm 0.05$) continues down to at least log $N$(\CIV) =
11.75, a factor of 10 deeper than 
the previous determination at $z > 3$\ in Songaila (1997).  
By fitting down the Lyman series, we are able to determine accurate
\CIV/\HI\ ratios for 19 of the identified systems, rather than
relying on a statistical estimate of the median metallicity as has
often been done.  Dav\'{e} et al.\ (1998) have previously determined
that the scatter in \CIV/\HI\ in the Q1422+231 spectrum of Songaila \&
Cowie (1996) requires an intrinsic scatter in the [C/H] of a factor of
$\sim$ 3.  
Finally, by considering only the \CIV\ systems
above the detection limit, we obtain the same $\langle$log
$N$(\CIV)$\rangle$ as 
Dav\'{e} et al.\ (1998), a statistic used to infer that the mean [C/H] =
$-2.5$. However, we stress that these simulations do not take into
account the full complexity of the physical processes determining the
\CIV/\HI\ ratio in the IGM so that both the mean [C/H] and its scatter
may change with future more detailed work.

\section{Stacking}\label{stacksec}

As discussed in previous sections, the \CIV/\HI\ ratio in low $N$(\HI)
systems may hold the key to understanding metal enrichment mechanisms
in the IGM.
Since one of the major limitations in detecting weak \CIV\ lines is
S/N, stacking many sections of the spectrum is one way to circumvent
this problem (e.g.\ Norris, Peterson \& Hartwick 1983; Tytler et al.\ 1995).  

\begin{figure}
\centerline{\resizebox{0.85\textwidth}{!}{\includegraphics{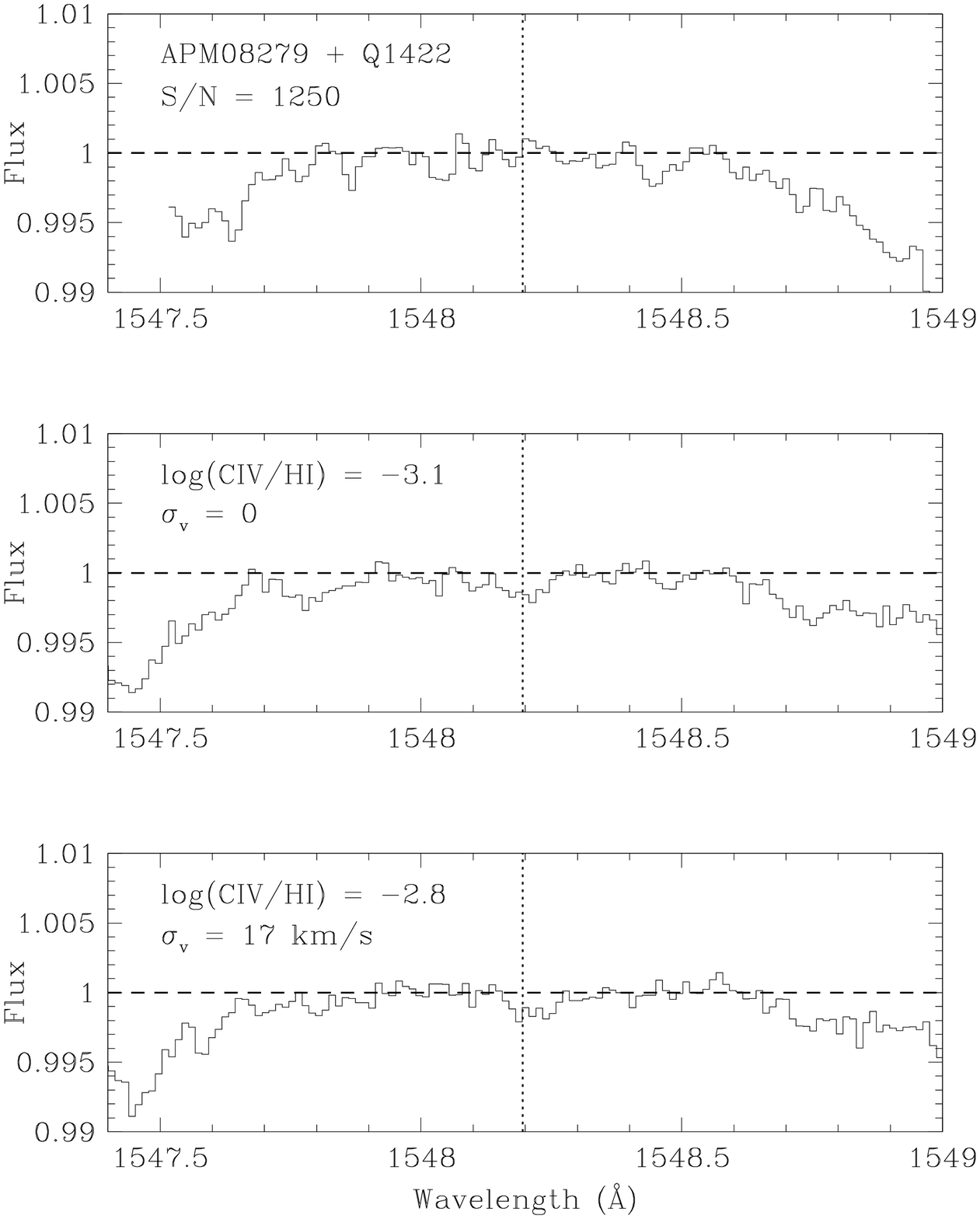}}}
\caption{Top:  The resultant composite stack having co-added a
total of 67 lines in 2 QSO sightlines, weighted by S/N.  Other panels:
Stacks produced from synthetic absorption spectra.  The \protect\lya\ forests
are reproduced from APM 08279+5255 and Q1422+231 and seeded with a
C\,IV/H\,I ratio and offset, $\sigma_v$ as indicated.  The $b$-values used
to simulate the C\,IV absorption lines have been taken to be $b$(C\,IV) =
$\frac{1}{2}$ $b$(H\,I).
The composite spectra shown here for the model spectra represent 4 $\sigma$
detections and have equivalent widths of $\sim$ 0.15 m\AA.}
\label{data_stack}
\end{figure}

In our application of this technique, we select \lya\ lines with 13.5 $<$ log
$N$(\HI) $<$ 14.0 whose corresponding \CIV\ spectrum shows no obvious
metal lines or contamination from other absorption features.  Each
section is de-redshifted to the rest frame, re-binned to the
dispersion of the lowest $z$ system, weighted according to its S/N,
stacked and finally re-normalized.  The optimal weighting used to
co-add the \CIV\ sections is given by

\begin{equation}
w_j = \frac{1/\sigma_j^2}{\sum_j 1/\sigma_j^2}
\end{equation}

where $\sigma_j^2$ is the variance of the data. To achieve maximum
sensitivity, we use low column density \HI\ lines 
from both the APM 08279+5255 and Q1422+231 spectra.  Within the ranges
defined in \S3, a total of 67 low $N$(\HI) lines were identified in 
the two QSO sightlines.
\textit{The resulting composite spectrum
has a S/N = 1250 and shows no absorption at the rest wavelength of 
\CIV, $\lambda_0 = 1548.2$}, as can be seen from the top panel of 
Figure \ref{data_stack}.  

In order to place a significance limit on this
non-detection, a synthetic spectrum was created, re-producing the
\lya\ forest of the 2 quasars from the fitted \HI\
line lists.  \CIV\ was included for \HI\ lines with log $N$(\HI) $<$
14.5 assuming a fixed \CIV/\HI\ ratio and $b$(\CIV) =
$\frac{1}{2} b$(\lya) (representing a combination of thermal and bulk
motion, as found in Paper I).
Figure \ref{data_stack} shows the results of stacking synthetic
spectra with log \CIV/\HI\ =
$-3.1$.  The resultant absorption in this stack
has an equivalent width of 0.15 m\AA\ and represents a 4$\sigma$
feature which we therefore adopt as the detection limit for our stacked data.
Several analyses (e.g.\ Songaila \& Cowie 1996; Paper I) have
determined the \CIV/\HI\ ratio in high column density \lya\ clouds to be
in the range from log \CIV/\HI\ = $-2.9$ to $-2.6$.  The detection limit
of the synthetic stack is almost factor of two lower than the
metal-poor limit of this range, and thus it would appear to indicate a drop in
the \CIV/\HI\ ratio at lower \HI\ column densities.
It must be understood, however, that this technique has several
potential problems which may compromise its
efficiency in detecting weak absorption features.  We now discuss
these problems in turn.

\begin{itemize}

\item  In order to co-add each section, the data must be re-binned,
    usually to the dispersion of the lowest redshift system.  This
    could smooth out a weak feature, although the scale of smoothing
    is very small compared with the width of the expected line 
    so that this is not likely to be a major effect. 

\item  Even for a fixed carbon abundance there will be a scatter
    in the values of \CIV/\HI\ (as discussed in the previous section).  
    Therefore, if
    absorption is detected in the composite spectrum it will be
    averaged over a range of column densities that can not be
    recovered individually.  Moreover, depending on the scatter
    of \CIV/\HI\ values, the absorption may be dominated by the
    strong tail end of this metallicity range.  In fact, it is conceivable
    that residual absorption could be caused by only a few relatively
    strong lines, since this method relies on an average which is very
    sensitive to a non-Gaussian tail.  Interpreting a residual signal
    in the composite spectrum is therefore not straightforward,
    although in the analysis presented here we find that there is no
    absorption in the stacked data and so we can determine a
    useful detection limit.  

\end{itemize}

\noindent More serious problems for the present analysis are:

\begin{itemize}

\item  The sensitivity of this method to errors in continuum fitting, 
anomalous pixels and other forms of contamination.  The usual procedure is
    to visually inspect each section before adding it to the stack in
    order to ensure that it is `clean'.  This will filter out major
    contamination by, for example, uncorrected cosmic ray events or
    absorption due to systems other than \CIV.  However,
    small errors in the continuum fit or deviant pixels could
    seriously compromise the efficiency of the stack. In addition, a
    re-normalization  of the composite spectrum is usually required, since the
    small errors in the original continuum fit have now been
    compounded by stacking.  

\item Stacking the individual sections of spectrum in order to build up
    a signal is pivotal upon centering the absorption feature in the
    composite.  If there is a significant error in the stack center,
    i.e.\ if an  offset exists between
    the redshift of the parent \lya\ line and its associated \CIV\
    complex, the composite signal would be smeared out.  Depending on
    the magnitude of the offset, this effect may lead
    us to under-estimate the amount of absorbing material.  This
    problem could be exacerbated by the afore-mentioned need for a
    re-normalization because it may be difficult, if not impossible, to
    distinguish a weak smeared signal from the compound continuum errors.

\end{itemize}

\subsection{Redshift Offset Between \CIV\ and \lya}

\begin{figure}
\centerline{\resizebox{0.85\textwidth}{!}{\includegraphics{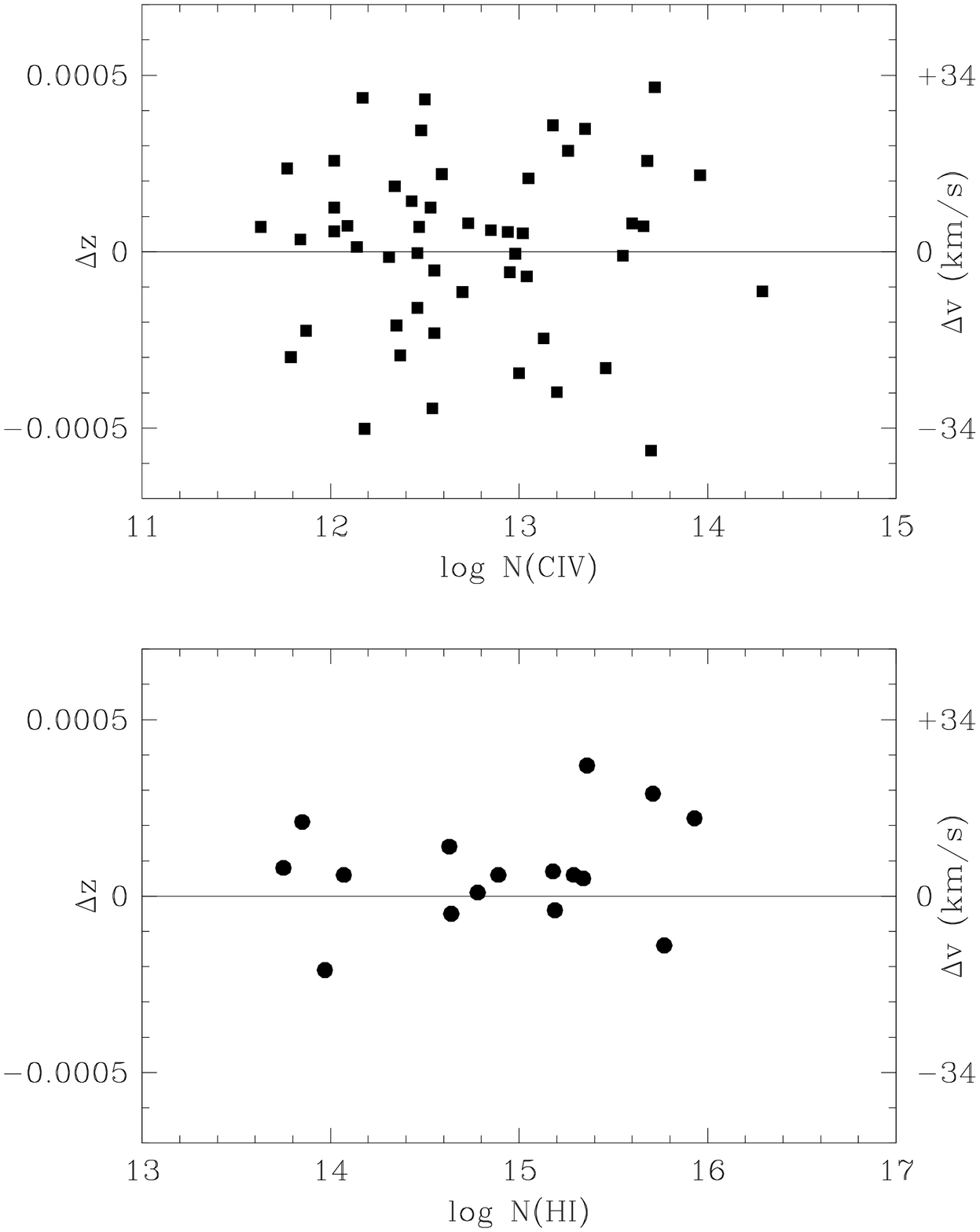}}}
\caption{Top panel:  Offset distribution calculated from the C\,IV centroids and
Voigt profile fitted redshifts of H\,I in the metal systems in APM
08279+5255 and Q1422+231.  Bottom panel:  For 19 C\,IV systems in Q1422+231,
an accurate $N$(H\,I) can be determined by tracing down the Lyman series.
In 16 of these systems, the C\,IV is associated with a single H\,I line,
the offsets for which are plotted here.  For the remaining 15 C\,IV systems,
higher order lines were either not available or
were too severely blended for line fitting.}
\label{offsets}
\end{figure}

In Paper I, we showed that there is indeed a random redshift
offset ($\Delta z$) 
between the measured position of the \lya\ and corresponding
\CIV\ lines.  In that work it was found that the redshift offset 
had a distribution with $\sigma_z \sim$ 4 $\times 10^{-4}$
equivalent to a velocity difference of 27 \kms, although
this statistic was based on a relatively small number of systems.  The
addition of a second quasar to the analysis has improved the statistics
and for a total of 56 \CIV\ systems in the two quasars we now determine a
$\sigma_z$=2.6 $\times 10^{-4}$  ($\sigma_v \sim$17 \kms\ for
an average redshift of 3.45).  
This redshift offset was determined in two different
ways.  In both cases the redshift of the \CIV\ was estimated by taking
the centroid of the system since it is often complex, consisting
of several blended
components.  For most of the corresponding \lya\ lines, the 
absorption is well represented by a single component and so in the first 
instance the redshift of the \HI\ cloud was obtained from the Voigt
profile fit. For comparison, the redshift of \lya\ was also determined
using the same centroid method applied to the \CIV\ systems and it was
found that both determinations yielded almost identical results for the
distribution of $\Delta z$, shown in
Figure \ref{offsets}.   Since it is the strongest component
in each \CIV\ complex that first emerges from the noise in the weak
systems, this will be the feature enhanced by the stacking procedure.
Therefore, we also investigated the distribution of offsets between the
fitted $z$ of \lya\ and the redshift of the deepest \CIV\ trough and
once again the value $\sigma_v \sim$  17 \kms\ was found.  

We investigated how seriously the composite spectrum would be
affected by this offset which effectively shifts 
the expected positions of the \CIV\ lines by a random amount.
Again, synthetic spectra were produced, this time including a redshift
offset applied to the position of the \CIV\ line with $\Delta z$ drawn
at random from a Gaussian distribution with $\sigma_z$=2.6$\times
10^{-4}$.  We find that in order to reproduce a 4$\sigma$ detection in
the presence of a redshift offset, twice as much \CIV\ must be included
in the \lya\ forest clouds, i.e.\ log (\CIV/\HI) = $-2.8$,
see the bottom panel of Figure \ref{data_stack}.
This is consistent with the measured \CIV/\HI\ in log $N$(\HI)$>$14.5
lines and is therefore not a sufficiently sensitive limit to establish whether
the low column density \lya\ clouds contain significantly less \CIV\
that the stronger lines.  We conclude that a factor of at least two
improvement in sensitivity is required in order to show
conclusively whether the low column density absorbers are more metal
deficient than their high column density counterparts.
Alternatively, it must be shown that the redshift
offset in these lines is $\ll$ 17 \kms\ and dilution of absorption
from smearing the stack is unimportant.

\begin{figure}
\centerline{\resizebox{0.85\textwidth}{!}{\includegraphics{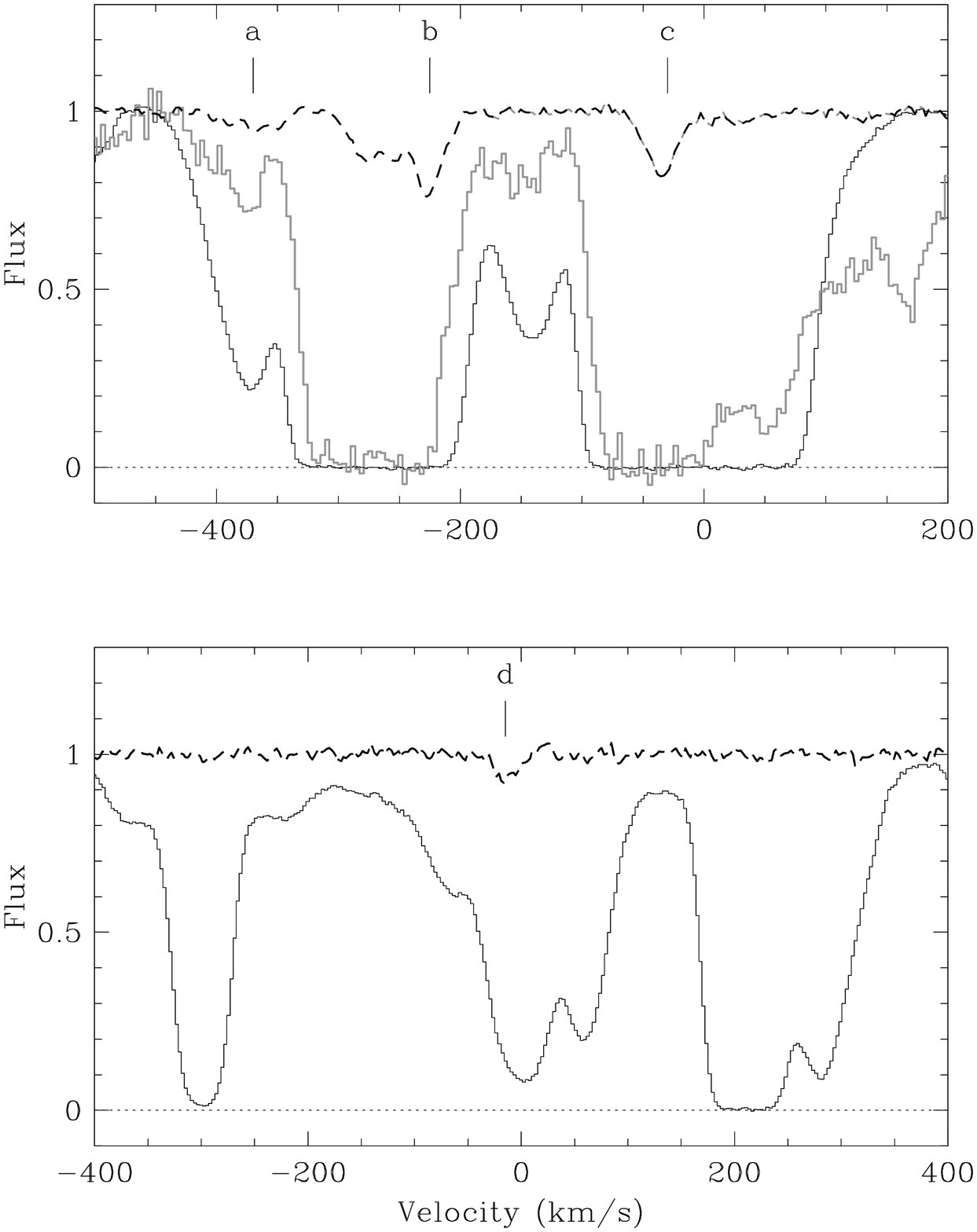}}}
\caption{Top panel:  Example of absorption components in H\,I and C\,IV
$\lambda 1548$ in Q1422+231.
The solid black line is \protect\lya\ absorption and the grey line is the
corresponding Ly$\beta$.  The dashed line depicts the C\,IV absorption
for 3 associated C\,IV complexes, marked a, b and c. The weak C\,IV system
`a' is associated with an unsaturated \protect\lya\ cloud and has a $\Delta
z < $ 1 \kms\  Bottom panel: A second example of an unsaturated \protect\lya\
line, here $\Delta z \sim$ 15 \kms.}
\label{lya_breakdown}
\end{figure}

The redshift offset in saturated lines could have two possible
explanations, one physical and the other an observational artefact.  
There could be an intrinsic redshift difference between
the \CIV\ and \lya\ absorbers, caused by, for example, ionization
effects or outflows.  An additional effect could be a redshift offset
caused 
by the blending of strong \lya\ lines which, when saturated, can not be
distinguished into separate components.  Two examples of this are
shown in Figure
\ref{lya_breakdown}. Plotted in velocity space are two examples of
\lya\ absorbers (black solid line) and corresponding \CIV\ (dashed
line).  Both saturated (systems `b' and `c') and weak (systems `a' and
`d') \lya\ clouds are shown and in the top panel the \lyb\ (in gray)
is also included.  The \CIV\ system `c' is associated with an
apparently monolithic \lya\ absorber centered at $v$ = 0 and exhibits a
redshift offset of approximately 25 \kms.  However, as seen from the
\lyb\ absorption, this system is made up of more than one component
and the \CIV\ clearly has a redshift that more closely matches the
strongest of these.  The unsaturated \lya\ absorber associated with the
\CIV\ system `a' does not break down into a multi-component system in
\lyb\ and the redshift offset is correspondingly small ($\Delta z <$
1 \kms).  However, there are other examples of \CIV\ systems associated
with weak \lya\ that \textit{do} show a significant offset, for example system
`d' in the lower panel of Figure \ref{lya_breakdown}.  Unfortunately,
higher order Lyman lines are not available for this particular case.
We re-measured the redshift offset for each of the 19 \CIV\ systems in
Table \ref{hifits} for which an accurate $N$(\HI) had been obtained by
tracing down the Lyman series.  For 16 of these systems, the \CIV\
appeared to be associated with a single \HI\ component (i.e.\ not
obviously blended), as seen from \lyb\ and/or \lyg.  The offsets
associated with the sub-cloud whose redshift most closely matches that
of the \CIV\ are shown in the lower panel of Figure \ref{offsets}.
Given the small 
number of systems for which the Lyman series can be traced, at least
to \lyb, the statistics of the offset distribution are not very
meaningful.  However, the scatter is now clearly smaller, although 
the high $N$(\HI) \lya\ lines at the top end of
the column density distribution may still have several sub-components
that are unresolved in our data and the offset may be further reduced
if the Lyman series could be traced down to subsequent transitions.  

\subsection{An Improvement on the Stacking Method?}

\begin{figure}
\centerline{\resizebox{15cm}{!}{\includegraphics{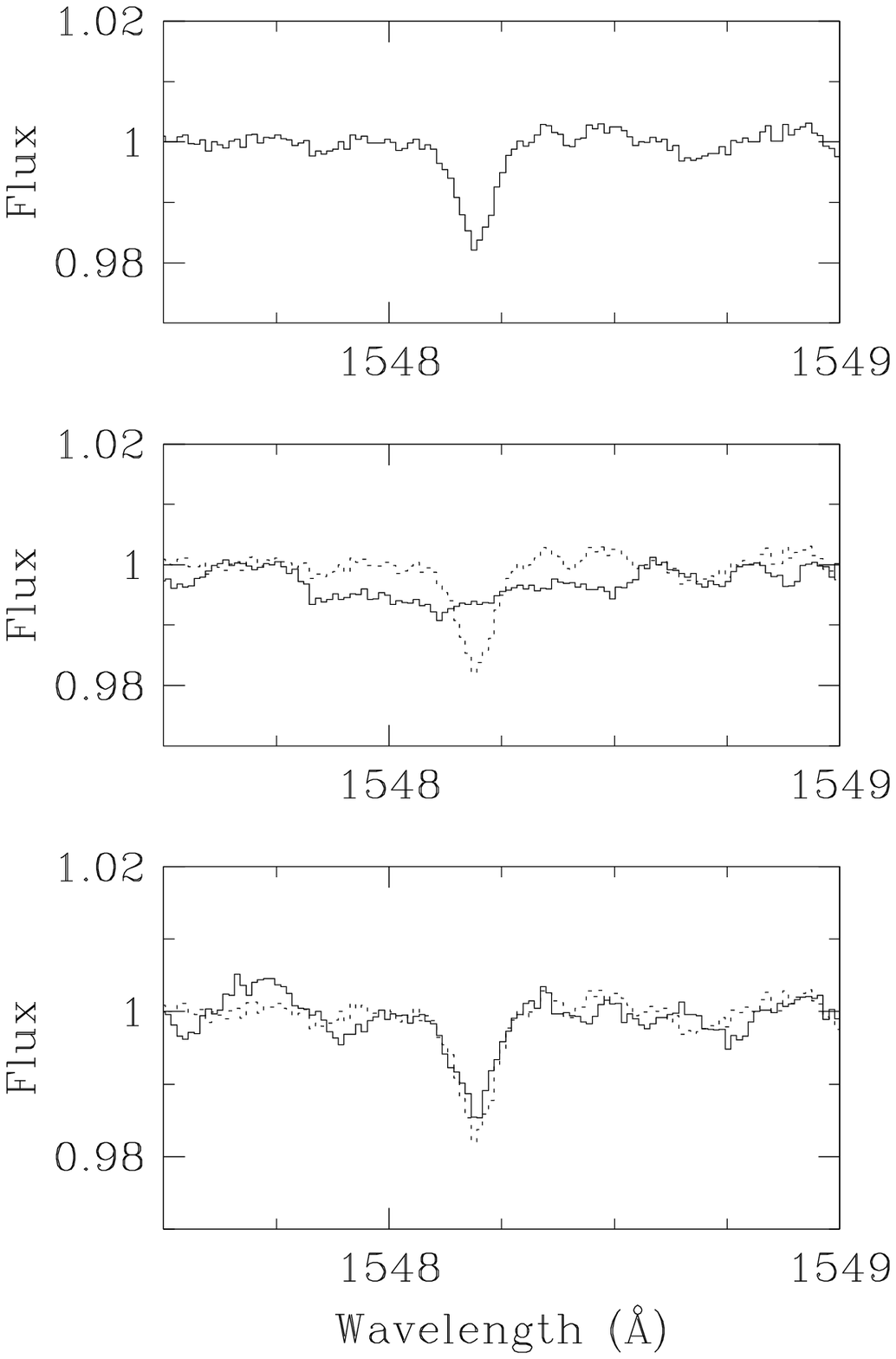}}}
\caption{An illustration of how re-centering absorption lines can
eliminate the effect of an offset.  Top panel:  Synthetic stack
produced by simulating a \protect\lya\ forest with S/N = 200 and log C\,IV/H\,I =
$-2.0$.  Composite stack produced for \protect\lya\ lines with 13.5 $<$ log
$N$(H\,I) $<$ 14.0.  Middle panel:  As above but with an added offset
($\sigma_v = $ 17 km/s).  The unshifted stack (from the top panel)
is dotted in for comparison.  Bottom panel:  As in the middle panel,
but with the spectra re-centered on the $\tau_{max}$ pixel before
stacking.  Comparison with the unshifted spectrum
(dotted profile) shows that almost all of the original signal is
recovered.}
\label{preshift}
\end{figure}

We investigated a possible solution to the problem of an unknown
offset from the predicted position of \CIV\ systems  
discussed above.  The technique involves scanning 
for the maximum absorption around the predicted position of \CIV\ 
absorption and re-centering the stack at this wavelength. In practice,
each (de-redshifted) data section is scanned $\pm 2 \sigma_z$ 
from the predicted \CIV\ line center (i.e.\ 
for $\Delta z$ = 0) and the section re-centered on the 
pixel with the maximum optical depth ($\tau_{max}$) prior to stacking.
Figure \ref{preshift} shows an example of how this 
technique is clearly able to improve upon a direct stack in the
presence of a $\Delta z$ for a synthetic spectrum containing lines with
log \CIV/\HI\ =$-$2.0.  It can be seen that without re-centering, 
whilst the overall level of the continuum is below unity for the
`smeared' stack, all profile information is lost. Such a broad
depression may be mistaken for compound errors in the continuum level
and lost in the subsequent renormalization of the stacked spectrum.
Executing a
re-center on the $\tau_{max}$ pixel before stacking recovers
almost all of the original signal in this simulation, the
characteristic line profile is prominent and will not be lost when the
post-stacking continuum is re-fitted.  

\begin{figure}
\centerline{\resizebox{0.85\textwidth}{!}{\includegraphics{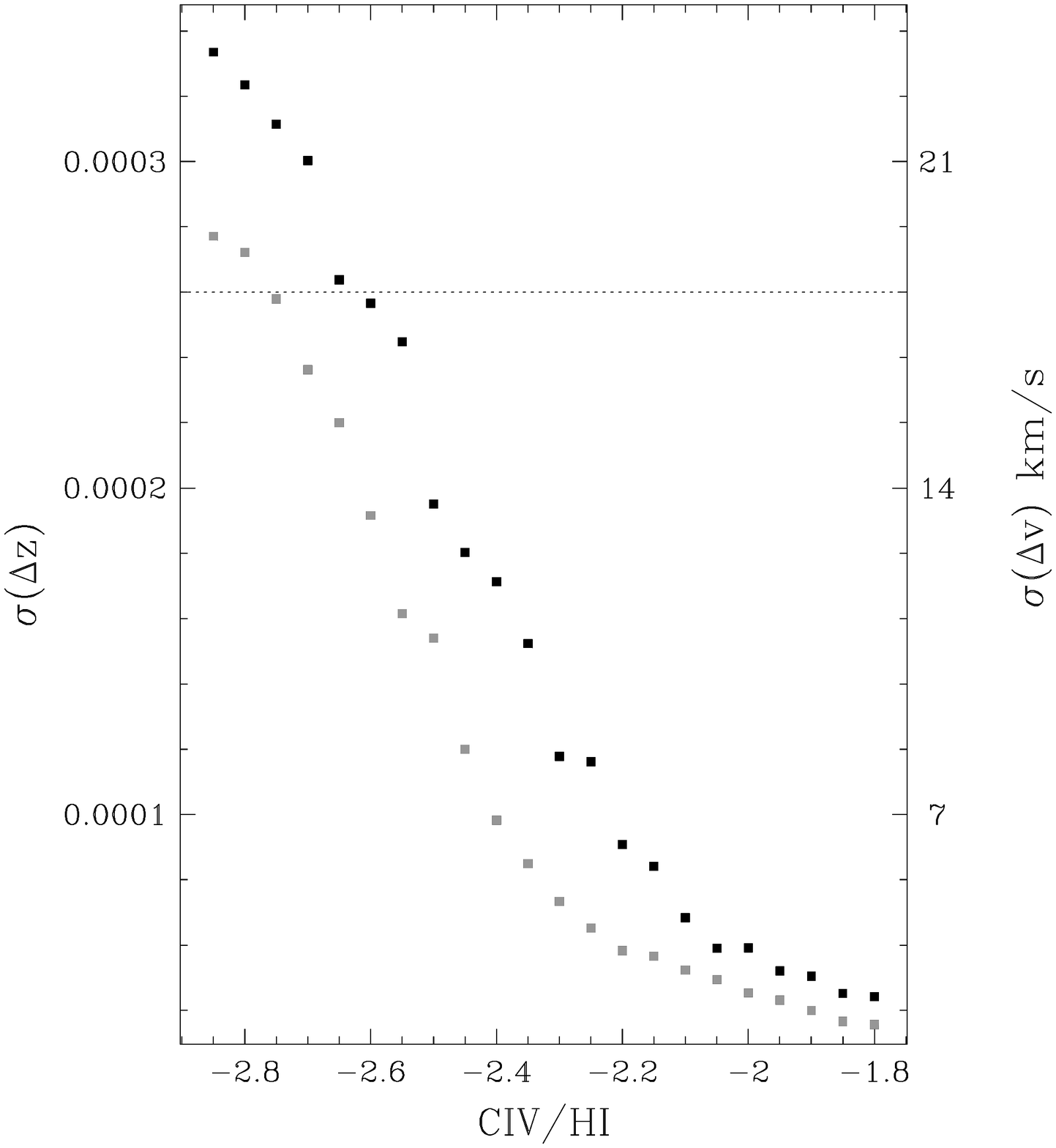}}}
\caption{Results from a pre-shift feasibility study. For each C\,IV/H\,I
ratio, 1000 C\,IV absorption lines are produced with $b$=13 km/s and S/N =
200, corresponding to log $N$(H\,I) = 13.75.  For each line, the offset
between the true line center and the position of $\tau_{max}$ is
calculated.  The black squares represent the $\sigma$ of the 
offset distribution for each
simulation of 1000 lines and is plotted in terms of both redshift and the
equivalent velocity (assuming an average $z=3.45$).  The gray squares
are the results of the same procedure but with a 3 pixel smooth
applied to the spectrum prior to locating $\tau_{max}$.  The measured value
of the offset determined from 55 C\,IV systems in APM 08279+5255 and
Q1422+231 ($\sigma_z = 2.6 \times 10^{-4}$ or $\sigma_v =$ 17 \kms)
is marked in the a dotted line.  }
\label{test_shift}
\end{figure}

However, as one attempts
to detect progressively weaker absorption, the likelihood of centering on
a noise pixel becomes higher and this technique is no longer
efficient.  In order to determine whether a `re-center' is a viable
improvement to the stacking method, given the S/N of the data and the low
column density of the targeted features, we performed a feasibility study
in which test absorption lines were created over a range of
\CIV/\HI\ ratios.  The results of this study are shown in Figure
\ref{test_shift} where we plot the distribution of $\sigma_z$ caused
by line center misidentification (i.e.\ where $\tau_{max}$ is a noise
feature rather than line trough) as a function of \CIV/\HI\ ratio.
Clearly, when the $\sigma$ of the offset caused by trough
misidentification exceeds the value of the intrinsic offset we are
trying to overcome, this re-centering technique is no longer useful. 
However, these are conservative estimates for how well the centering 
would perform on real data, since we would expect some trough
misidentification from contaminating (i.e.\ anything but \CIV) lines in
addition to the effects of noise investigated here. 
Since we need to determine whether the low column density
\lya\ lines have the same metallicity as their high $N$(\HI)
counterparts, we must ideally reach levels of sensitivity deeper than
log \CIV/\HI=$-2.9$. Figure
\ref{test_shift} shows that for log \CIV/\HI\ $< -$2.6, this technique no
longer compensates adequately for the intrinsic offset and is
therefore unable to improve the stacking technique in the search for
weak lines.  Smoothing the spectrum over several pixels
before locating $\tau_{max}$ improves the pre-shifting slightly as
shown by the gray squares in Figure \ref{test_shift}.  However, even
with smoothing, this re-centering procedure
can not compensate for a $\sigma_z$ = 2.6
$\times 10^{-4}$ below a log \CIV/\HI\ = $-2.75$.

\medskip

In summary, by stacking together the \CIV\ regions associated with low
column density \lya\ lines in Q1422+231 and APM 08279+5255, we have
obtained a S/N = 1250 composite spectrum which shows no residual \CIV\
absorption.  Using simulated spectra we find that for log \CIV/\HI\ =
$-3.1$ in the weak \lya\ lines we would expect a 4$\sigma$ detection 
of the composite absorption.  The shortcomings of this technique are 
discussed and we focus on an observed redshift offset between the 
position of \lya\ and its associated \CIV.  The redshift offset
determined from the detected \CIV\ systems has a dispersion $\sigma_z$
= 2.6$\times10^{-4}$ which corresponds to a $\sigma_v$ = 17 \kms.  
We have investigated how this offset would 
affect the stacking procedure by using simulated spectra and find that
a $\sigma_v$ = 17 \kms\ will reduce the sensitivity of this method by
a factor of two and that a metallicity of log \CIV/\HI\ = $-2.8$ is now
required to achieve a 4$\sigma$ detection.  We are unable to know
whether the same redshift offset persists in the low $N$(\HI) clouds 
targeted by
the stacking technique, but a large range of offsets are
measured over the full column density range of detected \CIV\ systems,
so one must clearly take into account the possible repercussions when
interpreting the stacked data.  
 
\section{Analysis of Optical Depths}\label{tausec}

A potentially more robust way of measuring weak absorption in high
S/N spectra is to analyse the optical depths in each \lya\ forest
pixel and its corresponding \CIV\ pixel (Cowie \& Songaila 1998).  
This technique also has the advantage that by tracing down the Lyman
series, one can determine the \CIV/\HI\ ratio in each 
pixel over a very large range of $\tau$(\lya).  In Paper I we
investigated the potential of this method with APM 08279+5255 and
critically analysed its performance on synthetic
spectra.  Specifically, we investigated the effect of including a
redshift offset and how the results depended on the $b$-value of \CIV.
We concluded that,  
despite the excellent quality of the spectrum, 
the results from the spectrum of APM 08279+5255 were inconclusive and
could not 
determine whether the metallicity of the IGM
is constant or diminished at low values of \HI, because at the lowest values of
$\tau$(\lya) both scenarios were consistent with the observations.
The spectrum of Q1422+231 presented here not only has a significantly
higher S/N than that of APM 08279+5255, but has many of its higher
order Lyman lines accessible for analysis and considerably less contaminating
absorption by, for example,  \MgII\ systems (Ellison et al.\ 1999b).  
This spectrum therefore represents
the best data yet obtained for this analysis of the low column density
\lya\ forest.

\subsection{The Analysis Procedure}

Briefly, the optical depth technique consists of stepping
through the \lya\ forest measuring $\tau$(\lya) for each pixel.  
The noise ($\sigma$) array is used to determine which pixels are
included in the 
analysis via a series of optical depth criteria to account
for effects such as saturation.  Using the values in the noise
arrays rather than fixing the rejection criteria provides maximum
flexibility for this technique so that it can be readily applied to
spectra of different S/N ratios.

For pixels with a residual flux ($F = e^{-\tau}$ for a normalised
spectrum) $F < 3 \sigma$ above 
the zero level, we trace down the Lyman series since  there is too
little residual flux in these saturated pixels to determine an
accurate optical depth from \lya\ alone.  
However, the danger here is that higher order
lines may be contaminated by lower redshift \lya.   We therefore use
$\tau$(\lya) = 
minimum($\tau$(Ly$n$)$f_{\rm{Ly\alpha}}\lambda_{\rm{Ly\alpha}}$ /
$f_{\rm{Ly}n}\lambda_{\rm{Ly}n}$) over all observed higher order lines
using the higher order pixel if $3 \sigma < F$(Ly$n) < (1 - 3
\sigma)$.  This minimizes the effect of contamination 
and maximizes the number of usable pixels and range of $\tau$(\lya) 
which can be considered for analysis.  If no $3 \sigma < F$(Ly$n) <
(1 - 3\sigma)$ pixels are found, then the
pixel is discarded.  
The position of the associated \CIV\ $\lambda\lambda$1548, 1550 lines
are calculated and, again to avoid contamination, we use $\tau$(1548)=
minimum($\tau(1548)$,$2\tau$(1550))\footnote{The ratio of the line 
$f$-values (oscillator strengths) 
in the \CIV\ doublet is 2:1} if the flux in the 1550 \AA\ component $<
1 - 3 \sigma$, otherwise only $\tau$(1548) is considered.    The \CIV\
optical depths are 
then binned according to their corresponding $\tau$(\lya) and the
median determined for each interval.
Taking the median of a large number of pixel optical
depths not only provides a statistical advantage over considering a
relatively small number of lines (as for the stacking method), but
also is much less susceptible to 
non-Gaussian effects.  In order to estimate 1 $\sigma$ errors for the optical
depth determinations, we used bootstrap re-sampling with 2\AA\
sections of the \lya\ forest and corresponding \CIV, i.e.\ we drew $n$
random sections from the complete set of $n$ sections (in this case $n
= 361$) that comprise
the original data, with replacement.
This procedure was repeated 250 times and the 1$\sigma$ error was
taken to be the dispersion of these 250 realizations.

\subsection{Results}

\begin{figure}
\resizebox{\textwidth}{!}{\includegraphics{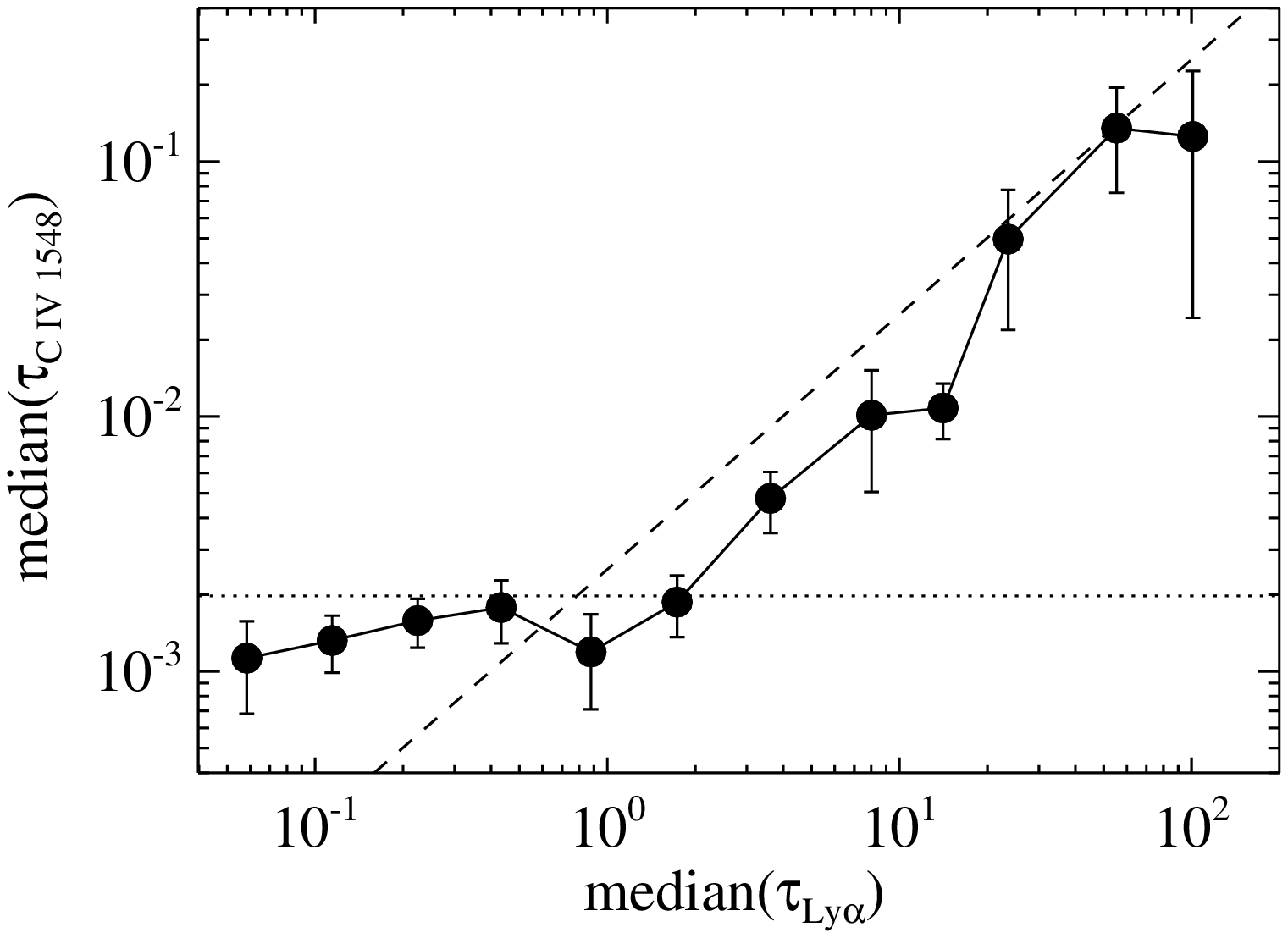}}
\caption{Results from the optical depth analyses of Q1422+231.  Error
bars are 1$\sigma$ as determined from the bootstrap method, as
described in the text.  The median absorption determined from pixel
pairs separated by the C\,IV doublet wavelength ratio over the entire
wavelength range considered  
(redward of \protect\lya) is shown by the dotted horizontal line.
The dashed diagonal line shows the optical 
depths expected from a spectrum with a constant log C\,IV/H\,I = $-2.6$. }
\label{1422_tau}
\end{figure}

Figure \ref{1422_tau} shows the results obtained for the optical depth
analysis of Q1422+231.  The shape of the optical depth
distribution appears consistent
with a constant level of \CIV/\HI\ (i.e.\ parallel to the dashed line)
over optical depths from $\tau$(\lya) $\sim$ 100 down to $\sim$ 2 --
3, below which $\tau$(\CIV) flattens off to an approximately constant
value.  Each of the low optical depth bins ($\tau$(\lya) $<$ 3) contains
approximately 10\% of the \HI\ pixels.   This percentage decreases with
increasing $\tau$(\lya) with only approximately 2\% of pixels in the
$\tau$(\lya) = 10 bin.
In addition, we determine optical depths in pixel pairs separated by 
the \CIV$\lambda\lambda$1548, 1550 doublet ratio \textit{regardless of
$\tau$(\lya)} over the entire range considered for \CIV\ absorption.
This is done using the same method of doublet comparison as before to
eliminate contamination.   
The median of this reference distribution is plotted in Figure
\ref{1422_tau} as a dotted line and represents 
the median absorption for an effectively random set of pixel pairs
separated by \CIV\ doublet ratio
and will include the effects of noise and low level contamination
expected to affect our results.  This median optical depth is
higher than the observed $\tau$(\CIV) for all bins with
$\tau$(\lya)$\simlt$ 1.  This is a significant result which indicates
that the \CIV\ absorption for these
optical depths is less than that expected by selecting
random pixel pairs.  The key question here is whether the
signal in the low optical depth \HI\ pixels is due to \CIV\ absorption
in low density regions of the IGM or whether the flattening of the
data is caused by some 
limiting factor in our analysis such as contamination or noise.

\subsection{Testing the Results}

\begin{figure}
\resizebox{15cm}{20cm}{\includegraphics{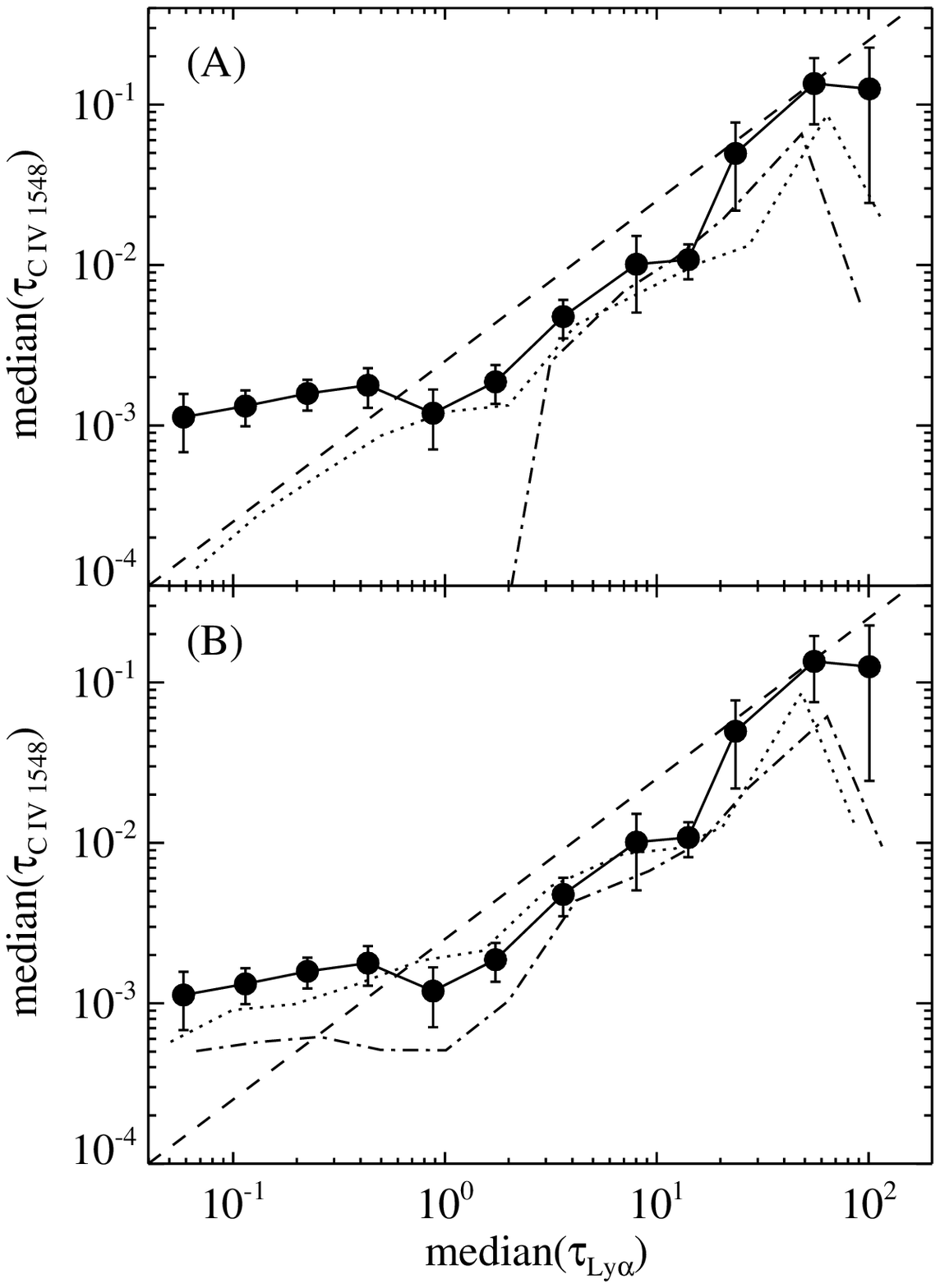}}
\caption{The optical depth distribution of Q1422+231 (solid points) is
shown compared with the results from simulated spectra.
Panel `A': The dotted line shows the optical depth analysis of a
simulated spectrum with constant log C\,IV/H\,I = $-2.6$ in all \protect\lya\
lines and the dot-dash line
shows the results from simulated spectrum which has log
C\,IV/H\,I = $-2.6$ only in log $N$(H\,I) $>$ 14.5 lines, with no noise
added to either spectrum.  Bottom panel: As for panel `A' except that
a S/N ratio (taken from the error array) has now been added 
to both spectra.  All four spectra use the real \protect\lya\
forest of Q1422+231 and have had C\,IV added to the synthetic spectrum
based on the fitted H\,I linelist.  In addition all four synthetic
spectra have a $\sigma_z$ = 2.6 $\times
10^{-4}$ and $b$(C\,IV) = $\frac{1}{2} b$(\protect\lya).   The dashed line
indicates constant log C\,IV/H\,I = $-2.6$.}
\label{2panel}
\end{figure}

As we saw in the previous section, thorough simulations of the analysis
technique are vital for interpreting our results. 
Here, we have taken the \lya\ forest directly from the data and
artificially enriched it with \CIV\ to test our methods and address several
questions.  In Paper I we explored the effect of $b$-values,
redshift offset and noise on synthetic spectra in order to determine
whether a break in the \CIV/\HI\ distribution could be distinguished
from a constant ratio in the spectrum of APM 08279+5255.  To review
these findings and to give a visual impression of the optical depth
analysis, we show the results of this technique on 
four synthetic spectra in Figure \ref{2panel} and compare them with the
results from Q1422+231 (solid points).  The top panel (`A')
of this figure shows the optical depth analysis of 2 spectra, one of which
has a constant log \CIV/\HI\ = $-2.6$ in all \lya\ lines (shown as a
dotted line) and a second spectrum which has log
\CIV/\HI\ = $-2.6$ only in log $N$(\HI) $>$ 14.5 lines (dot-dashed line),
with no noise 
added to either spectrum redward of \lya.  The bottom panel shows the same
spectra as panel `A' except that noise has now been added to both spectra,
based on the error array of the actual data (typically S/N = 200
redward of \lya).  All four spectra use the real \lya\ 
forest of Q1422+231 and have had \CIV\ added to the synthetic spectrum
based on the fitted \HI\ linelist.  In addition, all four synthetic
spectra in Figure \ref{2panel} have a $\Delta z$ = 2.6 $\times
10^{-4}$ and $b$(\CIV) = $\frac{1}{2} b$(\lya).   The dashed line
indicates constant log \CIV/\HI\ = $-2.6$.  

There are several
points to note here.  Firstly, in the absence of noise, there is a
clear drop in $\tau$(\CIV) below $\tau$(\lya) $\sim 3$ if no \CIV\
is added in \lya\ lines with log $N$(\HI) $<$ 14.5.  However, this
steep decline is much less drastic when noise is included, and
although $\tau$(\CIV) shows a steady decrease down to $\tau$(\lya)
$\sim$ 1, below this value it flattens off to an approximately
constant value.  The inclusion of noise also causes the same apparent
flattening in the spectrum of constant \CIV/\HI, although the
$\tau$(\CIV) in this spectrum is consistently above the value measured
for the dot-dashed line.   There is also a small contribution to this
flattening from line blending which changes the overall slope of the
expected \CIV/\HI, an effect that is exacerbated for larger $b$(\CIV).  
  This shows that even in relatively
high S/N spectra, distinguishing a break in the \CIV/\HI\ distribution
is very difficult.  Instead, as we shall see later in this section,
one of the main uses of this method is to determine whether the
optical depths measured with this technique can be adequately
accounted for with the detected \CIV\ systems or whether there must be
significant amount of \CIV\ still below our current detection limit.   
We also note
that for large $\tau$(\lya) the measured 
$\tau$(\CIV) is always less than expected from the dashed line, given
the input ratio of 
\CIV/\HI.  This is due to two effects.  Firstly, since this technique
considers $\tau$(\lya) as the minimum value obtained by tracing down
the Lyman series when a line is saturated, if contamination is
successfully removed we will tend to
under-estimate $\tau$(\lya) at these optical depths due to noise
(remember that the real \lya\ forest is used in the simulated spectra
so that this will be an effect even in panel `A' of Figure
\ref{2panel}).  Secondly, the \CIV\ that is included in the synthetic
spectra is based on a linelist of fitted \HI\ values which will
probably not be accurate for saturated lines.  These effects also
account for the drop of $\tau$(\CIV) at very large $\tau$(\lya) which
contain $<$ 1\% of the total \HI\ pixels and are therefore very uncertain.

Clearly, the exact results of the optical depth analysis are sensitive to
the combination of several factors including blending, $b$-values and
noise, which we will discuss further later in this 
section.  Therefore, rather than attempting to directly fit the
observed distribution of optical depths in Q1422+231 we aim to determine
whether the optical depths
measured in the spectrum can be explained solely by the relatively
strong \CIV\ absorbers detected directly and presented in \S4 or whether
the results from this analysis are indicative of additional \CIV.  If
the latter is true, are these metals in the low density IGM?
We must also investigate whether our results 
can be explained in terms of contamination, noise or some other
limiting factor in the data.  Finally, we consider the effect of 
scatter and redshift offset ($\Delta z$), which have been found 
to compromise the efficiency of the stacking method. 

\begin{figure}
\centerline{\resizebox{10cm}{20cm}{\includegraphics{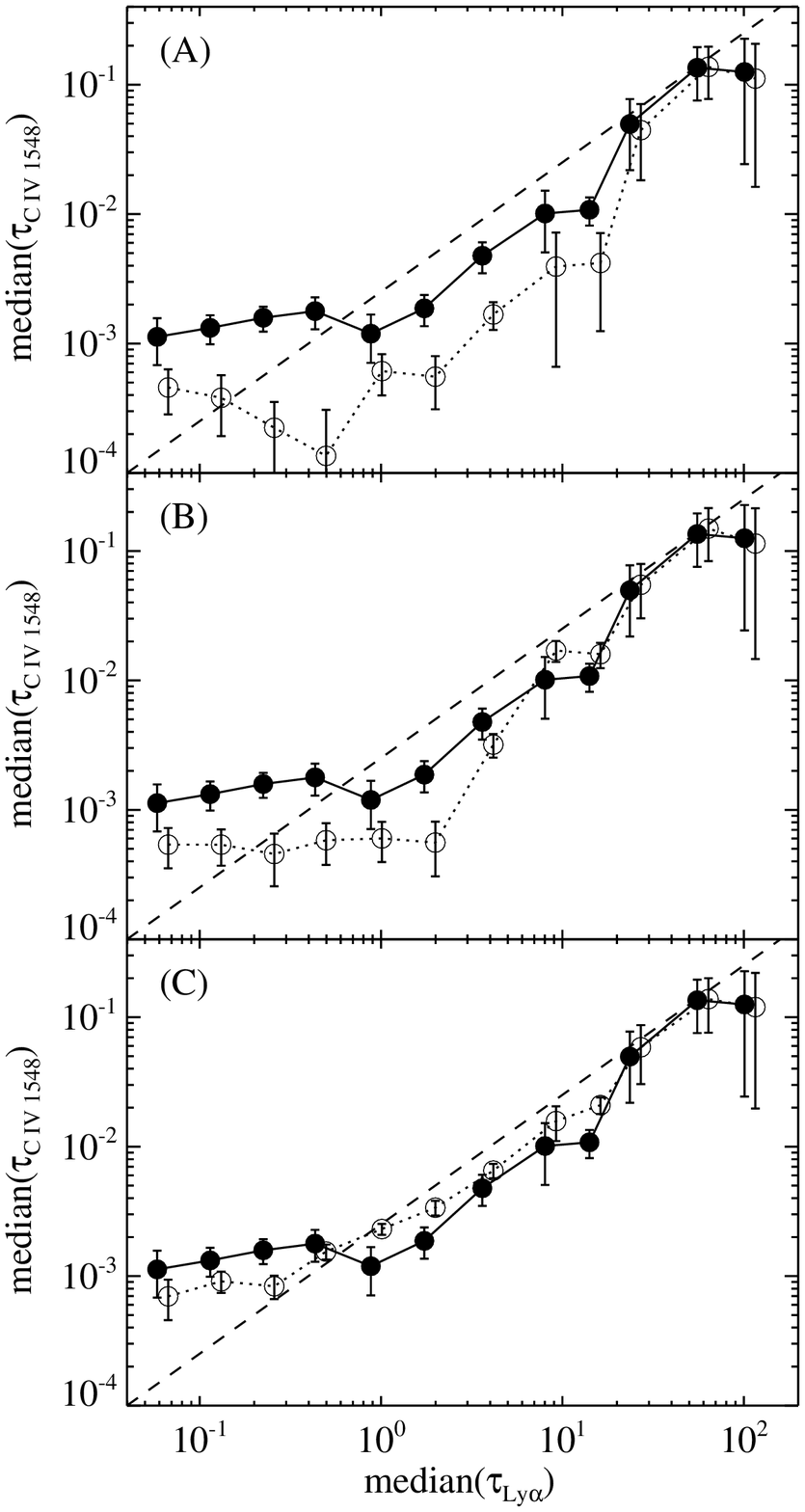}}}
\caption{The results from the optical depth analysis of Q1422+231
(solid points) are compared with three synthetic spectra.  Top panel:
open circles show the measured optical depths in a synthetic spectrum
enriched solely with the detected C\,IV systems listed in Table
\ref{civlist}.  Middle panel: In addition to the detected C\,IV
systems, log $N$(C\,IV)= 12.0 is included in all \protect\lya\ clouds with log
$N$(H\,I) $> 14.5$.  This spectrum therefore represents the maximum
amount of metals that could be `hidden' below the detection limit in
strong absorbers.  Bottom panel:  Supplementary C\,IV is now added in
all weak (log $N$(H\,I)$<14.5$) \protect\lya\ lines with log C\,IV/H\,I =
$-2.6$.  Clearly, more C\,IV is present than currently identified
directly and these optical depth results show that the data are
consistent with a significant amount of C\,IV in low column density clouds. }
\label{testabc}
\end{figure}

Three synthetic spectra were produced with \lya\ forest
absorption taken directly from the data (i.e.\ not reconstructed from a
linelist) and the 34 detected \CIV\ systems re-produced from the Voigt
profile parameters in Table \ref{civlist}. Spectrum `A'
has had no further metals added and therefore shows the results
expected of an optical depth analysis if we had already uncovered all
of the \CIV\ in the spectrum.  Spectra `B' and `C' have both 
been enriched with additional metals.  In spectrum `B', \CIV\ is
included in all strong (log $N$(\HI) $>$ 14.5) lines with $N$(\CIV) = 12.0,
i.e.\ the detection limit for directly identifiable \CIV\ systems.
This spectrum therefore represents the maximum amount of \CIV\ that
could be `hidden' in high column density \lya\ lines. In addition to 
this `hidden' \CIV\ in strong \HI\ lines and the fitted \CIV\ in Table 
\ref{civlist}, spectrum `C' has been enriched with a constant log 
\CIV/\HI\ = $-2.6$ in weak \lya\ lines ($N$(\HI) $<$ 14.5).
The results from analysis of these three synthetic spectra
are compared with the data in Figure \ref{testabc}.
For all three spectra, we include noise taken from the $\sigma$ error
array, a random redshift 
offset in the position of \CIV\ ($\sigma_z = 2.6 \times 10^{-4}$) and
take $b$(\CIV) = $\frac{1}{2}$ $b$(\lya).

The conclusions that can be drawn from Figure \ref{testabc} are as
follows.  First, there is clearly more \CIV\ in the data than we
have directly identified in \S4, since the dotted line of the
synthetic spectrum in the top panel is well below the solid 
line at all but the very highest $\tau$(\lya) points.  For
$\tau$(\lya) $\gs$ 3, the \CIV\ optical depths can
be recovered when additional metals are added into strong \lya\
absorbers, but below our current detection limit (spectrum `B').  This
supports the results of \S4 where we determine that
$f(N)$ is consistent with a power law distribution that
continues down to
log $N$(\CIV) = 11.75, with no evidence for a turnover in column
density distribution.  However,
adding \CIV\ at the limit of our detection in log $N$(\HI)$>$ 14.5
lines with no extra metals in weaker \lya\ clouds can not reproduce
the measured $\tau$(\CIV) in $\tau$(\lya)$<$ 3 pixels.  The results from
spectrum `C' in the bottom panel of Figure \ref{testabc} which include
log \CIV/\HI\ = $-2.6$ in weak \lya\ lines show that the observed \CIV\
optical depths in the low $\tau$(\lya) pixels are consistent with the
presence of some metals in significantly lower column density \HI\
clouds, although for $\tau$(\lya) $>$ 1 $\tau$(\CIV) exceeds the
observed value.  From Figure \ref{2panel} we have also seen that a
constant log \CIV/\HI\ = $-2.6$ in all \lya\ lines is a good
approximation to the data.  We stress here that these simulations are not an
attempt to fit the observed distribution of optical depths, rather
they are tests to determine whether the \CIV\ systems in Table
\ref{civlist} can account for the measured absorption.  If not, then
the objective is to investigate in which \HI\ column density regime
additional metals could be added in order to achieve the observed
quota.   

The results from these simulations show that the analysis of optical
depths in the spectrum of Q1422+231 is consistent with hitherto
undetected \CIV\ in both the strong and weak \lya\ lines.  Whilst the
results from our determination of the column density distribution from
detected \CIV\ systems is consistent with a power law function $f(N)$ that
continues down to at least log $N$(\CIV) = 11.7, these optical depth
results provide \textit{direct evidence} that there are more metals
lurking below our current detection limit.  

We investigate the effect that contaminating lines may have on this
result.  The analysis may be affected not only by low level
contamination from other metal lines such as \SiIV\ or telluric
absorption (strong absorbers will have been rejected by the \CIV\
doublet strength comparison discussed in \S6.1), but also effects
due the non-uniformity of ionizing sources and noise/fluctuations
in the continuum level due to fitting errors.  This latter effect
will cause fluctuations around the true continuum which will
be important only if they are greater than the level of the noise.  An
additional (possibly systematic) error may be present if the low order
polynomial fit to the continuum is consistently over or
under-estimated.  To test the
continuum fit of the \CIV\ regions, sections of the data redward of \lya\ 
deemed to contain no obvious
absorption were examined.  From a total of $\sim$ 5000 pixels, a mean
flux of 1.0002 and a median of 1.0001 were determined, vaules one order of
magnitude lower than $\tau$(\CIV) $\sim 10^{-3}$.  It therefore seems
unlikely that a systematic error in the continuum fit is
the cause of the observed flattening of \CIV\ optical depths seen in
Figure \ref{1422_tau}.  
The distribution of noise pixels would also suggest that there is 
no significant effect from weak telluric lines.
Errors in the continuum fit blueward
of \lya\ emission, however, are likely to be a more serious effect,
firstly because selection of absorption-free zones is difficult  and
secondly because the S/N is lower (50 -- 150).  The result of
continuum fitting errors in the forest in this analysis
will be to classify pixels with small $\tau$(\lya) into the wrong 
optical depth bin.  This will effectively associate \CIV\ with the
wrong $\tau$(\lya) and averaging out the metal absorption by this
`mis-binning' could explain the observed flattening of the
measured optical depths.  The effect of a small continuum error is
therefore similar to the effect of noise, affecting only those
pixels with optical depths smaller than the fitting
error.  We also find that
including a large number of weak contaminating lines such as \SiIV\ or \MgII\
could increase the measured $\tau$(\CIV) at low $\tau$(\lya) to the
constant level determined.  However, whilst realistic column densities 
do have a small effect on the optical depth result, it is unlikely that
weak contaminating lines account for all of the observed absorption.  

\begin{figure}
\resizebox{\textwidth}{!}{\includegraphics{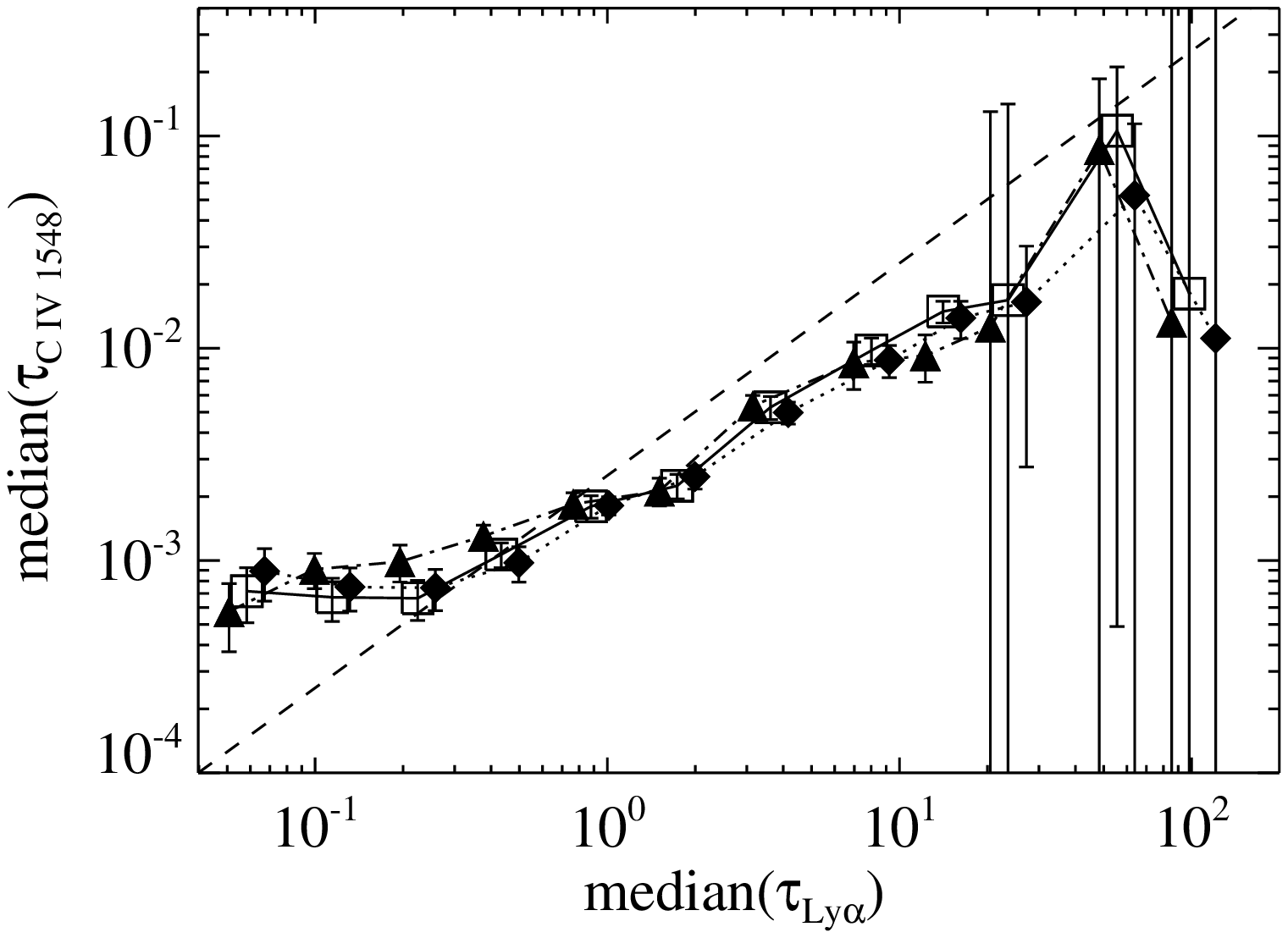}}
\caption{The optical depth results from three synthetic spectra with 
the \protect\lya\ forest re-created from
the spectrum of Q1422+231.  In all cases $b$(C\,IV) = $\frac{1}{2}
b$(H\,I).  The open squares (joined with the solid line) show the
optical depths measured for an input log C\,IV/H\,I = $-2.6$
(with no scatter and no $\Delta z$) in \protect\lya\ lines of all H\,I column
densities.   The solid diamonds (connected 
with a dotted line) have C\,IV/H\,I = $-2.6$, but include a scatter
of $\sigma = 1 \times 10^{-3}$.  The spectrum represented by the
solid triangles (dot-dash line) has no scatter but a redshift offset
with  $\sigma_z
= 2.6 \times 10^{-4}$ has been included in the position of the C\,IV
lines.  Bootstrap 1$\sigma$ errors from 250 iterations are shown and
points have been offset here to facilitate distinction.  
Clearly the effect of scatter and redshift offset on the
results of this analysis are minimal.
}
\label{scatter}
\end{figure}

As discussed previously, a scatter is expected in the observed 
\CIV/\HI\ values even if [C/H] and the ionizing background remain uniform.  
To investigate the importance of this effect, we simulated a `control'
spectrum by re-producing directly the \lya\
forest of Q1422+231 and using $b$(\CIV) = $\frac{1}{2} b$(\lya) and an
input log \CIV/\HI\ = $-2.6$ with no scatter and no redshift
offsets.  For comparison, we then produced a second spectrum
in the same way, but rather than adopting a constant \CIV/\HI,
a Gaussian scatter was introduced with
$\sigma$(\CIV/\HI) = $1 \times 10^{-3}$.   
The comparison between the two simulated spectra is presented
in Figure \ref{scatter} where the `control' spectrum 
is shown as open squares connected with a solid line and the spectrum
containing a scatter of metallicities is shown as solid diamonds
connected with a dotted line.  The points have been offset from one
another in the figure for clarity.  The points at high $\tau$(\lya)
are again very uncertain due to the reasons previously discussed with
regards to Figure \ref{2panel}. 
The error bars are estimated using the same bootstrap technique as
employed for the data. The results 
from the 2 spectra are entirely consistent with one another and
indicate that a Gaussian scatter will therefore
not affect the overall median of $\tau$(\CIV).  Whilst the extent and
magnitude of the scatter in the distribution of \CIV/\HI\ values has
not yet been fully investigated in simulations
down to column densities below log $N$(\HI) $\sim$ 14.0 --
13.5, this result should not be significantly affected unless the
scatter becomes highly non-Gaussian at low $N$(\HI).  Also
plotted in Figure \ref{scatter} are the results of the optical depth
analysis if a redshift offset is included in the position of the \CIV\
line.  As before, metals are added to all \lya\ lines 
with a metallicity log \CIV/\HI\ = $-2.6$.  There is no scatter in
these values, but an offset in the position of \CIV\ has been included,
drawn at random from a Gaussian distribution with $\sigma_z = 2.6
\times 10^{-4}$ ($\sigma_v$ = 17 \kms).
These points are also consistent with the control spectrum, so we can
conclude from this simulation batch that neither scatter in the
metallicity nor redshift offset will have a large effect on the
outcome of our optical depth analysis.

Finally we note that more spectra are required in order to provide a
representative analysis of the high redshift IGM, since our view of
the enrichment of the \lya\ forest probed with a single sightline
is clearly blinkered.  From the list of \CIV\ systems 
in Table \ref{civlist} it can be seen that these absorbers are not
uniformly distributed in redshift.  For example, splitting this
spectrum in two by redshift produces vastly different optical depth
distributions due to a relative dearth of \CIV\ systems that spans over 300 \AA\
in this spectrum in the range $3.13 < z_abs < 3.33$.    With more
spectra, there is the 
possibility of not only developing a more representative study of these
high redshift absorbers but also investigating the redshift evolution of \CIV\
systems.  

\section{Conclusions}

In this paper, we have addressed the enrichment history of the IGM by
studying the \lya\ forest and its associated \CIV\ systems in
a very high S/N ($\sim 200)$, high resolution ($\sim$ 8 \kms) spectrum
of the well-known lensed quasar, Q1422+231 obtained with Keck/HIRES.  
The numerous \CIV\ systems associated with
high column density \lya\ absorbers are fitted with Voigt profiles
defined by a redshift, $b$-value and column density for each component
line.    We investigate the \CIV\ column density distribution,
$f(N)$, to very sensitive levels and detect several weak \CIV\ systems which
had not been previously identified in lower quality spectra of the same
object.  We determine a power law index $\alpha = 1.44 \pm 0.05$ which
continues down to log $N$(\CIV) = 12.2 before starting to turnover. By
simulating synthetic absorption lines with $b$-values taken at random
from the observed distribution, we estimate a correction factor to
account for incompleteness and find that the corrected data points 
now indicate that the power law continues down to at least log
$N$(\CIV) = 11.75, a factor of ten more sensitive than 
previous measurements (e.g.\ Songaila 1997). 
This shows that even at these low column densities there
is no evidence for a flattening of the power law and
therefore there are probably many more \CIV\ systems 
that lie below the current detection limit. 

We investigate two methods with which it may be possible to recover these
weak \CIV\ systems.  Firstly, we select 67 \lya\
lines with 13.5 $<$ $N$(\HI) $<$ 14.0 in Q1422+231 and APM 08279+5255 
and produce a stacked spectrum centered on the predicted position of
\CIV\ $\lambda$1548.  The composite stack has a S/N = 1250 and shows no
residual absorption; we use synthetic stacked spectra to determine
a 4$\sigma$ upper limit of log \CIV/\HI\ = $-3.1$.  
We critically assess the accuracy of this method
by performing the stacking procedure on a suite of simulated,
synthetic spectra and identify several associated problems.  We
investigate, in particular, the effect of a redshift offset between
the position of the \lya\ line and its associated \CIV.
With improved statistics, we refine the redshift offset determined in
Paper I and find that a $\sigma_v$=17 \kms\ is present in the \CIV\
systems which we detect directly.  By including a random redshift
offset drawn from a Gaussian distribution with $\sigma_v$=17 \kms\
in our stacking simulations, we find that log \CIV/\HI\ = $-2.8$ is now required
to achieve a 4$\sigma$ detection.  This limit is still consistent with
current measured metallicities in higher column density \lya\ clouds
and is therefore not sufficiently sensitive to determine whether the
\CIV/\HI\ ratio drops in low $N$(\HI) lines. A feasibility study is 
performed to assess the effectiveness of a `re-center' on the maximum 
optical depth pixel prior to stacking for removing
the effect of an unknown offset.  We find that this technique
can not improve the quality of the stack result in the \CIV/\HI\ regime 
that we are targeting.  It is not yet clear whether the observed
redshift offset persists in the low column density \lya\ clouds, but
it must be considered a factor.  Moreover, the effects of
contamination, continuum fitting errors and anomolous pixels also pose
problems for this technique, although in simulations the redshift
offset appears to be a major effect.  Therefore, if this
technique is to be pursued further to reach a meaningful detection
limit, an improvement in S/N by at least a factor of two is required. 

The optical depth technique introduced by Cowie \& Songaila (1998) is
considered as an alternative approach.  This technique is shown to
exhibit several advantages over the stacking method such as its
insensitivity to redshift offsets and its ability to exclude
contamination from other absorption features.  We develop this
technique as a method that
can be used to test whether the detected \CIV\ systems represent the
full tally of absorbers.  The data have optical depths
consistent with an almost constant log \CIV/\HI\ $\sim -3$ down to
$\tau$(\lya)$\sim$ 
2--3, below which $\tau$(\CIV) flattens off to an approximately
constant value.   It is unlikely that this flattening is real and is
most probably caused by the effect of noise and/or continuum errors,
even at such high S/N 
ratios as have been achieved in this spectrum.  Given the many effects
that may alter the measured $\tau$ values, such as blending and noise,
we do not 
attempt to fit the observed distribution of optical depths.  Instead,
our strategy is to test whether the detected \CIV\ systems are
sufficient to reproduce the measured $\tau$(\CIV) and if not, determine
how much additonal \CIV\ may be present below our current detection
limit.   By simulating
synthetic spectra with different enrichment recipes, we have shown
that the \CIV\ systems detected directly in the spectrum are not
sufficient to reproduce the results of the optical depth analysis of
Q1422+231.  
This is in agreement with the conclusions drawn from the column density
distribution of \CIV, i.e.\ that the data are consistent with a
continuous power law 
$f(N)$ down to at least $N$(\CIV) = 11.75 and that there is therefore likely to
be a large number of weak metal lines not yet directly detected. 
This agrees with the conclusions of Cowie \& Songaila (1998).

In order to interpret the results from the optical depth method, we have
simulated synthetic spectra with a range of input \CIV/\HI\ ratios.
We find that including \CIV\ associated with strong 
\lya\ lines ($N$(\HI) $>$14.5) but below the current detection limit,
in addition to the 34 identified \CIV\ systems,
can reproduce the optical depths measured in the observed spectrum for
$\tau$(\lya) $>$ 3.  For smaller values of $\tau$(\lya), some
additional metals are required and we find that including \CIV\ in low
column density \HI\ lines ($N$(\HI) $<$ 14.5) with log \CIV/\HI\ $=
-2.6$ produces optical depth results consistent with those
measured in the data.  However,
determining the precise \CIV/\HI\ in the low $N$(\HI) \lya\ clouds and
the density to which the enrichment persists is still uncertain due to
effects such as noise and continuum fluctuations and it is therefore
not possible to say whether the low column density forest is pristine.
Nevertheless, 
we find that even in the high
optical depth \HI\ pixels (which will not be seriously affected by
noise or small
continuum errors) the detected \CIV\ systems are not sufficient to
cause all the measured absorption and clearly there are more metals in
the IGM than we can currently detect. 

\acknowledgements

The authors would like to thank Bob Carswell, Alberto Fernandez-Soto, 
Matthias Steinmetz and the anonymous referee for discussions and
useful suggestions.  We are grateful to Len Cowie for his support and
enthusiasm towards this project.   
SLE and JS are supported by a PPARC postgraduate award and JS 
acknowledges additional funding from the Isaac Newton Trust.
We are also grateful to the expert assistance of the staff at the Keck
I telescope for their help in obtaining the spectra presented here.

%
%

\begin{figure}
\epsscale{0.9}
\plotone{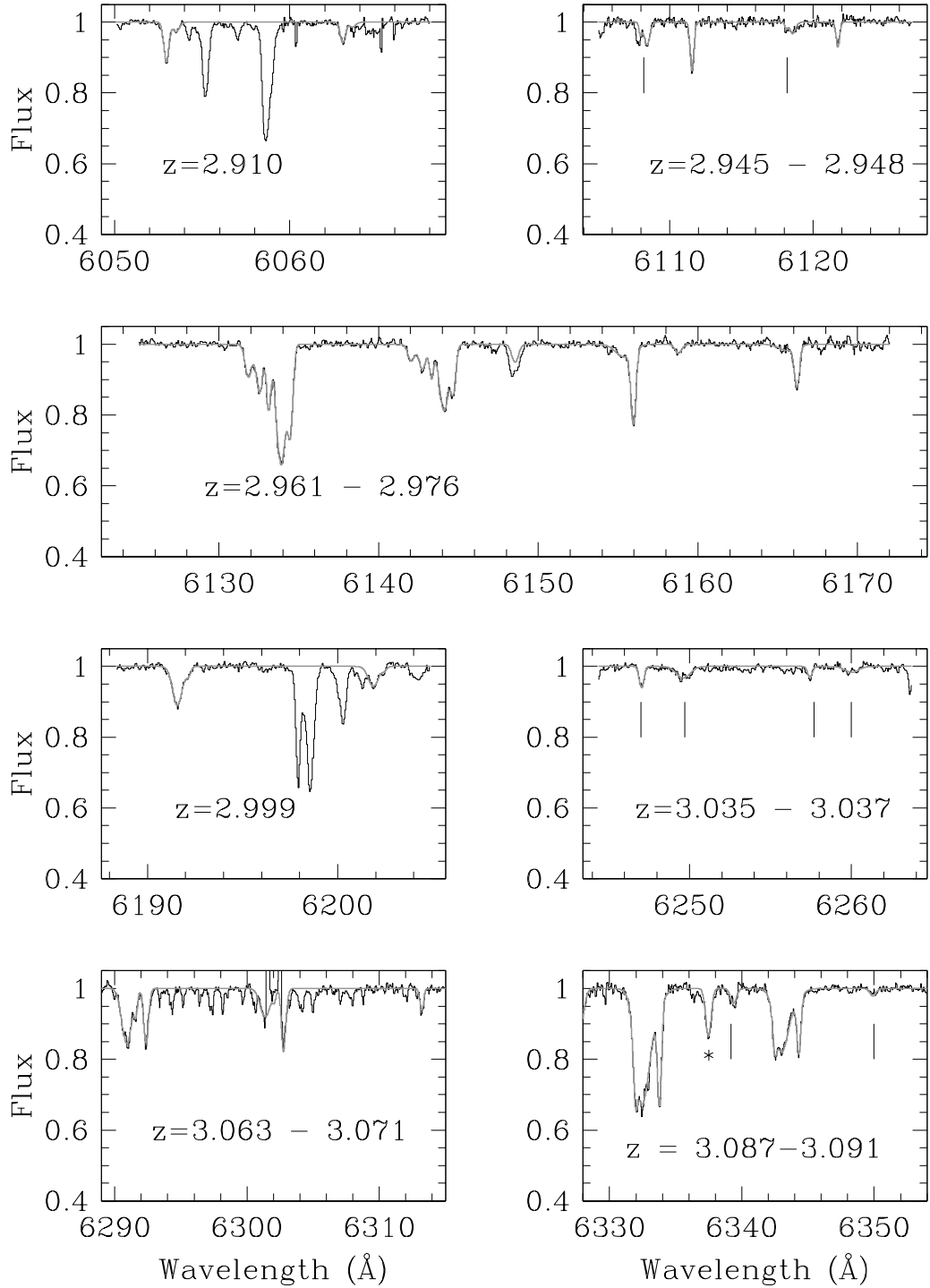}
\caption{Atlas of C\,IV doublets associated with \protect\lya\ lines in
our working region. The $x$-axis is wavelength in \AA\ and the $y
$-axis is normalized counts. Gray lines show the
profile fits with the parameters listed in Table \ref{civlist}.
The weakest C\,IV systems are indicated with tick marks to guide
the eye.  The system marked with a `*' in the bottom right hand panel
is a C\,IV $\lambda$ 1550 line whose corresponding 1548 \AA\ line is
blended with a lower redshift system and is not shown here.}
\label{civsys}
\end{figure}

\begin{figure}
\figurenum{1b}
\epsscale{0.9}
\plotone{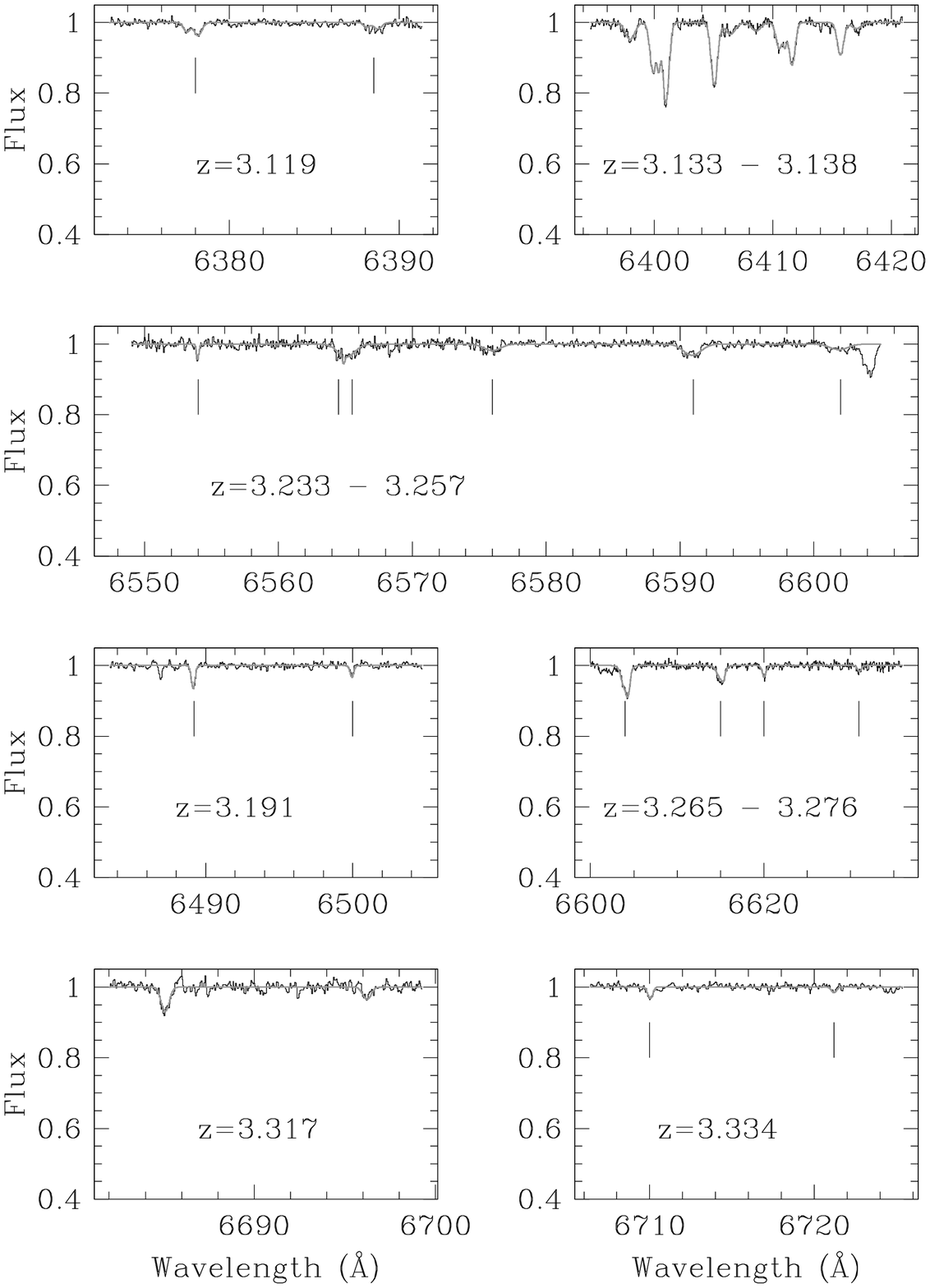}
\caption{Continued}
\end{figure}

\begin{figure}
\figurenum{1c}
\epsscale{0.9}
\plotone{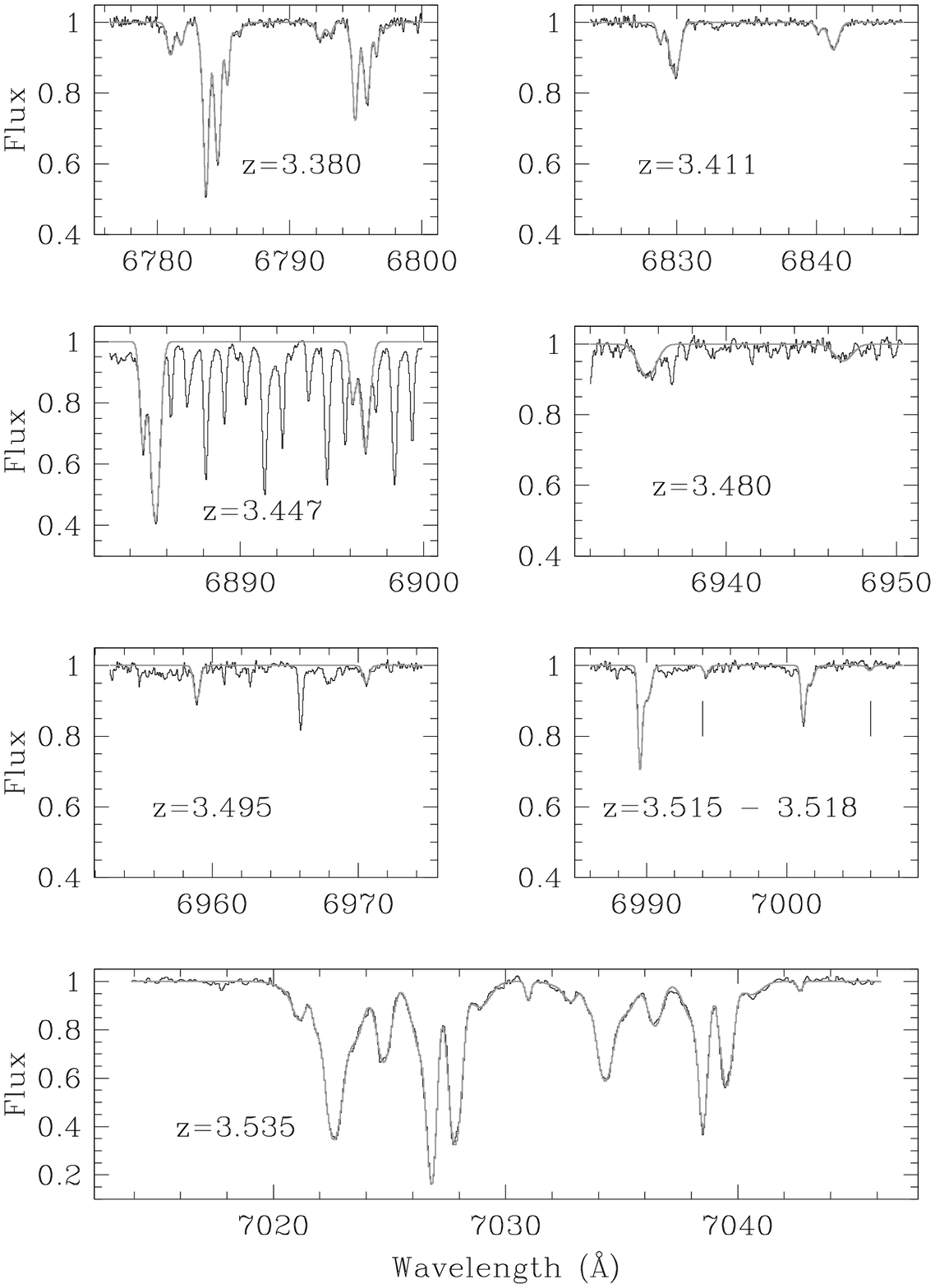}
\caption{Continued}
\end{figure}

\newpage

%
%

\begin{deluxetable}{cccr}
\tablefontsize{\footnotesize} \tablecaption{
Details of Absorption Components Fitted to Each C~\textsc{IV} System.}
\tablewidth{0pt}
\tablehead{
\colhead{System No.} &
\colhead{Redshift} &
\colhead{log~$N$(C~\textsc{IV})} &
\colhead{$b$ (km s$^{-1}$)} 
}
\startdata
C1  &   2.90969 & 12.44 &     8.8 \\ 
    &   2.91005 & 11.90 &    11.4 \\ 
C2  &   2.94522 & 11.74 &     4.2   \\ 
    &   2.94551 & 12.32 &    12.6 \\ 
C3  &   2.94754 & 12.47 &    7.2 \\ 
C4  &   2.96065 & 12.50 &    13.4 \\ 
    &   2.96110 & 12.68 &    13.6 \\ 
    &   2.96146 & 12.61 &     7.6 \\ 
    &   2.96197 & 13.28 &    19.3 \\ 
    &   2.96235 & 12.80 &     9.7 \\ 
C5  &   2.97143 & 12.23 &    13.3 \\ 
C6  &   2.97584 & 12.46 &    35.3 \\ 
    &   2.97622 & 12.76 &     9.2 \\ 
C7  &   2.99922 & 12.66 &   16.7 \\ 
    &   2.99959 & 11.61 &     6.2 \\ 
C8  &   3.03505 & 12.14 &     8.3 \\ 
C9  &   3.03672 & 12.35 &    28.4 \\ 
C10 &   3.06338  & 12.98  &   24.5 \\
    &   3.06433  & 12.66  &   11.0 \\
    &   3.06383  & 12.01  &    5.6 \\
C11 &   3.07101  & 12.43  &    7.9 \\
C12 &   3.08666  & 12.95  &  12.8 \\ 
C13 &   3.08990  & 12.59  &    7.9\\ 
    &   3.09020  & 13.33  &   30.5 \\ 
    &   3.09051  & 13.10  &   47.8 \\ 
    &   3.09108  & 12.91  &    8.8 \\ 
C14 &   3.09468  & 12.17  &   14.1 \\ 
C15 &   3.11930  & 11.87  &   11.1 \\ 
    &   3.11973  & 12.12  &   14.4 \\ 
C16 &   3.13252  & 12.46  &   26.0 \\  
C17 &   3.13379  & 12.80  &   17.5 \\ 
    &   3.13409  & 12.27  &    6.4 \\ 
    &   3.13448  & 13.00  &   16.2 \\ 
C18 &   3.13712  & 12.89  &   16.9 \\ 
    &   3.13799  & 12.33  &   27.5\\ 
C19 &   3.19143  & 12.09  &    6.5 \\ 
C20 &   3.23330  & 11.77  &    2.6 \\ 
C21 &   3.24047  & 12.53  &   40.4 \\ 
C22 &   3.25716  & 12.50  &   42.5 \\ 
C23 &   3.26564  & 12.65  &   26.9 \\ 
    &   3.26584  & 11.95  &    9.8 \\ 
C24 &   3.27596  & 12.02  &   13.0  \\ 
C25 &   3.31710  & 12.35  &   13.0  \\ 
C26 &   3.33410  & 11.84  &    7.7 \\ 
C27 &   3.37994  & 12.55  &   15.5 \\ 
    &  3.38045   & 12.30  &  12.3 \\ 
    &  3.38135   & 11.87  &   5.0 \\ 
    &  3.38167   & 13.27  &  11.6 \\ 
    &  3.38223   & 13.23  &  14.0 \\ 
    &  3.38271   & 12.61  &   8.1 \\ 
    &  3.38316   & 12.22  &  20.4 \\ 
C28 &   3.41080  & 12.20  &   10.7 \\ 
    &   3.41149  & 12.90  &   21.2 \\ 
C29 &   3.44691   & 12.97  &   8.5 \\ 
    &   3.44736   & 13.46  &  13.4 \\ 
C30 &   3.47963   & 12.85  &  31.7 \\ 
\enddata
\label{civlist}
\end{deluxetable}

\begin{deluxetable}{cccr}
\tablenum{1}
\tablefontsize{\footnotesize} \tablecaption{Continued.}
\tablewidth{0pt}
\tablehead{
\colhead{System No.} &
\colhead{Redshift} &
\colhead{log~$N$(C~\textsc{IV})} &
\colhead{$b$ (km s$^{-1}$)} 
}
\startdata
C31 &   3.49488   & 12.46  &  11.4 \\ 
C32 &   3.51465   & 12.83  &   7.5 \\ 
    &   3.51497   & 12.59  &  15.3 \\ 
C33 &   3.51770   & 11.87  &  10.2 \\  
C34 &   3.53490   & 12.57  &  34.7 \\ 
    &   3.53505   & 12.69  &  16.0 \\ 
    &   3.53549   & 12.43  &   7.7 \\ 
    &   3.53599   & 13.69  &  19.9 \\  
    &   3.53659   & 13.12  &  21.7 \\ 
    &   3.53738   & 13.22  &  17.2 \\ 
    &   3.53872   & 13.58  &   9.2 \\ 
    &   3.53937   & 13.58  &  14.1 \\ 
    &   3.54005   & 12.82  &  25.6 \\ 
    &   3.53848   & 13.33  &  25.0 \\ 
    &   3.54140   & 12.10  &   4.4 \\ 
\enddata
\end{deluxetable}

\begin{deluxetable}{cccc}
\tablefontsize{\footnotesize} \tablecaption{Column densities for C\,IV
enriched systems  
H\,I clouds whose Voigt profile fits could be accurately determined
from \protect\lya\ (if unsaturated), or
\lyb\ and \lyg.  The C\,IV column densities are determined to better
than 5\%.
} 
\tablewidth{0pt}
\tablehead{
\colhead{Redshift} &
\colhead{log~$N$(H\,\textsc{I})} &
\colhead{log~$N$(C\,\textsc{IV})} &
\colhead{log C\,\textsc{IV}/H~\textsc{I}} 
}
\startdata
2.947 & 15.18$\pm$0.06 & 12.47 & -2.71\\
2.999 & 15.77$\pm$0.05 & 12.70 & -3.07\\
3.035 & 14.64$\pm$0.07 & 12.14 & -2.50\\
3.037 & 14.63$\pm$0.03 & 12.34 & -2.29\\
3.063 & 15.36$\pm$0.03 & 13.18 & -2.18\\
3.071 & 13.85$\pm$0.01 & 12.43 & -1.42\\
3.132 & 13.74$\pm$0.10 & 12.46 & -1.28\\
3.134 & 15.71$\pm$0.03 & 13.26 & -2.45\\
3.137 & 15.93$\pm$0.09 & 13.00 & -2.93\\
3.191 & 14.89$\pm$0.20 & 12.09 & -2.80\\
3.276 & 14.07$\pm$0.01 & 12.02 & -2.05\\ 
3.318 & 13.97$\pm$0.10 & 12.35 & -1.62\\
3.334 & 14.78$\pm$0.06 & 11.84 & -2.94\\
3.382 & 16.70$\pm$0.40 & 13.68 & -3.02\\
3.411 & 15.19$\pm$0.02 & 12.98 & -2.21\\
3.479 & 15.29$\pm$0.02 & 12.85 & -2.44\\
3.515 & 15.34$\pm$0.20 & 13.02 & -2.32\\
3.518 & 13.29$\pm$0.03 & 11.87 & -1.42\\  
3.539 & 16.17$\pm$0.20 & 14.29 & -1.88\\
\enddata
\label{hifits}
\end{deluxetable}

\end{document}